\documentclass[11pt,twoside,a4paper]{article} 
\usepackage{times}
\usepackage[utf8]{inputenc} 
\usepackage{authblk}
\usepackage{hyperref}
\usepackage{graphicx}
\usepackage{xr}
\usepackage{hyperref}
\newcommand{\nat}{Nature}

\begin{document}

\title{Large scale analysis of violent death count in daily newspapers to quantify bias and  censorship.}%

\author[1,2,3]{M. Casolino  }
\affil[1]{INFN Structure of Roma Tor Vergata,   Rome, Italy }
\affil[2]{Riken, Computational Astrophysics lab, Wako, Japan}
\affil[3]{{Dipartimento di Fisica, Universit\`a di Roma Tor Vergata, Rome, Italy}  \\ email \href{mailto:casolino@roma2.infn.it}{casolino@roma2.infn.it} }
 
  \maketitle 
 
\begin{abstract} 
In this work we develop a series of techniques and tools to determine and quantify the presence of bias and censorship in newspapers. These algorithms are tested analyzing  the occurrence of keywords `killed' and `suicide'  ({\it `morti', `suicidio'} in Italian) and their  changes over time, gender  and reported location on the complete online archives (42 million records) of the major US newspaper ({\it  The New York Times}) and the three major Italian ones ({\it  Il Corriere della Sera, La Repubblica, La Stampa}).  

Using these tools, since the Italian language distinguishes between the female and male cases,  we find the presence of gender bias in all Italian newspapers, with reported single female deaths  to be about one-third of those involving single  men.  

Analyzing the historical trends, we show evidence of censorship in Italian newspapers both during World War 1  and during the Italian Fascist regime. Censorship in all countries during World Wars and in Italy during the Fascist period is a historically ascertained fact, but so far there was no estimate on the amount on censorship in newspaper reporting: in this work we estimate that about $75\%$ of domestic deaths and suicides were not reported.  This is also confirmed  by   statistical analysis of the distribution of the least significant digit of the number of reported deaths. 
 
We also find that the distribution function of the number of articles vs. the   number of deaths reported in articles follows a power law, which is broken (with fewer articles being written) when reporting on few  deaths occurring in foreign countries. The lack of articles  is found to   grow with geographical distance from the nation where the newspaper is being printed.  
 
 Whereas the assessment of the truth of a single article or the debunking of  what are now called `fake news' requires specific fact-checking and becomes more difficult as time goes by, these methods can be be used in historical analysis and to evaluate quantitatively the amount of bias and censorship present in other  printed or online publication and can thus  contribute to   quantitatively assess the freedom of the press in a given country. Furthermore, they can  be applied in wider contexts such as the evaluation of bias  toward specific ethnic groups or specific accidents. 
 
\end{abstract}
 
\providecommand{\keywords}[1]{\textbf{\textit{Index terms---}} #1}

\section{Introduction}

Violent death is a dramatic event that can be  objectively quantified  in terms of the number of lives lost. Words such as `killed', `dead', `casualty', `suicide' etc... can thus be used to assess the presence of bias or censorship in reporting  incidents involving specific locations, groups of people, or time.

Coverage of death in the mass media  has been studied for the  insight it offers on how societies perceive this dramatic event\cite{aries1975western}; at the same time, selective death reporting can be used to  shape and distort public opinion.
A  6-months period study\cite{combs}  of  1975 issues of  the {\it Register Guard} and the {\it Standard Times} showed  that newspapers tend to over-report more news-worthy  death causes and overlook others that are  considered less interesting by the public. The work considered  the number of occurrences, the number of deaths reported, and the text surface area of the articles dealing with death and compared them with the statistical occurrences, finding that all forms of disease were under-reported whereas violent or catastrophic events were overrepresented.  An even more biased selection along these lines was found analyzing the news selection by television and radio\cite{altheide1977creating}.  
Studies on  the presence of  geographical bias in reporting accidental deaths have been conducted manually on small data sets: 
a 51-issues study (two months)  of two German  ({\it Frankfurter Allgemeine Zeitung} and {\it S$\ddot{u}ddeutsche$ $Zeitung$}) and two Australian newspapers ({\it The Australian} and {\it The Sydney Morning Herald}) \cite{doi:10.1080/14616700801997281} 
have shown a weak relationship with social proximity in the number of articles  reporting deaths. Similarly, a 6-months study   of the German tabloid {\it Die Abendzeitung}\cite{doi:10.1177/0267323188003001005}    reported evidence of a correlation between the number of accidental  deaths and the length of the article as well as a negative correlation of  the  distance between the town where the newspaper was printed and the event reported. In this case, the analysis involved the study of the amount of printed space  occupied by an article reporting of a fatal event. The authors did not find any significant correlation between the area taken by the article and the number of deaths reported\cite{stigler1989} but showed an inverse relationship (linear in logarithmic scale) of the area of printed text normalized to the number of victims and the distance (in km) of the event   from Munich, the town where the  {\it Die Abendzeitung} was printed.  They also found a minimum threshold of victims - increasing with distance - needed to elicit a report in the newspaper. They thus likened the response of the paper to the logarithmic response between stimulus strength and response intensity in human perception\cite{dember}.

On the opposite,  \cite{tsang1984} analyzed the occurrence of pictures in {\it Time} and {\it Newsweek} magazines. The study involved ten issues in three years for each magazine, for a total  of 60 issues. The  work involved measuring the area occupied by the picture, determine the country of origin of the picture, and classifying its violent or non-violent nature. The study  found that -  although domestic pictures were more common than foreign ones ({\it Time=66.4$\%$}, {\it Newsweek=71.2\%}) -  images depicting death were printed more often  when the event took place abroad ({\it Time$_{foreign}$}: $31\%$; {\it Newsweek}$_{foreign}$:$25\%$) compared to when it occurred in domestic USA ({\it Time$_{domestic}$}: $12\%$; {\it Newsweek}$_{domestic}$: $13\%$).  

   A more recent  study\cite{doi:10.1080/13576270802383840} showed that direct pictographic representations of death are very rare in western newspapers, amounting  only to  $4.5\%$ of the articles   (997 stories in a two-month time frame, 357 with pictures).  

Most of these studies have  been performed by manual scanning of the newspapers or magazines, therefore restricting the scope of the possible analysis.  With the digitization of printed literature, it is possible to perform systematic analysis on the entire archives of newspapers, magazines, and books. 
A large-scale  analysis of the number  of occurrences of notable personalities in books  has shown evidence of  censorship of various individuals during the  Nazi regime\cite{ScienceMichel176}. The books have    been digitally converted via OCR (Optical Character Recognition)\cite{10.1145/1577802.1577804} as part of the Google effort to digitize books and the occurrence of individual names has been compared with the  lexicon of the various languages\cite{niyogi}.

In this work, we have developed several algorithms that allow  a  systematic  study of  the occurrence of the articles involving fatal accidents and their gender/location. This allows to  assess the distribution laws and determine the presence of bias and censorship in reporting. We have used daily newspapers since they have the advantage of  addressing 
wider    audiences with higher frequency  and  a longer temporal consistency, often covering events usually ignored by books.

We have applied  these algorithms on  the full historical archives of   the three major Italian newspapers: {\it  Il Corriere della Sera} (Milan, CDS, 1876-2017), {\it La Repubblica} (Rome, REP, 1985-2017), and {\it La Stampa} (Turin, STA, 1867-2005) and the major US newspaper:  {\it The New York Times} (NYT, 1955-2017),  searching for  the articles containing selected  keywords: {\it `killed'} ({\it `morti'} in the Italian language, the mixed-gender plural term, as well as the singular female and male gender, {\it `morta'} and {\it `morto'} respectively), {\it `suicide'} ({\it `suicidio'}). These words have been chosen for their high rate of occurrence (3.1 million articles out of 42 million comprising the newspaper archives) in respect to similar lemmas (dead, casualties...)  and correspondence of usage since both are used to indicate violent deaths. Most importantly, in the case of the plural form they contain the number of people involved. Therefore, in addition to a simple word count, it is also  possible to study the distribution of the  number of people killed, the location, and  the category of victims.

The rest of the paper is organized as follows: in the {\it Methods} section is present the description of the archives, the retrieval and data processing steps, and the various algorithms. In the {\it Results/Discussion} section are shown the results on the time behaviour, gender, censorship, the scaling law, and the last digit analysis. Perspectives and future work are discussed in the {\it Conclusion} section. 
 
\section{Methods}

 Figure \ref{flow1}  shows the steps taken to access, parse, and identify the articles, from the access of the online archives to the creation of the various data sets, one  for each newspaper   and lemma considered.  All these tasks are  accomplished using Python scripts. The access to the online archive of each newspaper takes a few days to complete over the full time span, since this is mediated  by a web-based interface that usually returns ten articles at a time.    
The storage and handling of the data sets are performed using  the Root\cite{BRUN199781} framework and require a few minutes to complete.
This C++ based environment was developed at CERN to deal efficiently with high volumes ($\simeq$ Pbyte) of  data produced  from accelerator- space- and ground/underground-based detectors. As such, it is especially suited to create the ntuples (Ttrees in Root nomenclature) containing the data extracted from the newspaper articles.     All the selections, histogramming, and fitting have also been performed in this environment.

A scheme of the various types of algorithms used to  analyze the data sets and the main information they provide is shown in  Figure    \ref{flow2x}.

\subsection{Newspaper Archives}

	\begin{table}[t]
	\small
\hspace{-4cm}
\begin{tabular}{ c c c c c c c c c l}
                 
 \hline
Newspaper 	& Date & $T$           & & &  $K$ 				          &  	$N  $  			     & $G $ & $N\:and\:G$  & $\: \: $Archive  \\
						& 		 & 	all entries& 	Suicide & Morta& 	killed  	& 	number		  &  location    & number and  	&	$\: \: $ 				\\
						& 		 & 						 & 	Suicidio & Morto &			  morti				  	& 			present  &   present   &   location 	&					\\
	\hline
	NYT & 1852 &1.51e7& 1.43e5 & & 4.74e5  &3.10e4  &2.58e5& 1.82e4 &  https://www.nytimes.com/search/\\
	& 2016 & & &  & & & & &  \\
\hline 
CDS & 1876 &8.04e6& 7.72e4 & 3.33e5  &  2.77e5  &2.53e4  &7.44e4& 1.10e4 & http://archivio.corriere.it/ \\ 
& 2017 & & &  6.5e5 & & & & &  \\
\hline
REP & 1985 &6.36e6& 3.68e4& 6.52e4  &  1.68e5  &3.70e4  &8.36e4 &2.20e4 & https://ricerca.repubblica.it/ \\
& 2017               & & &  1.5e5 & & & & &  \\
 \hline 
STA & 1867 & 1.28e7& 7.33e4 & 1.31e5  & 1.79e5  &1.86e4  &1.39e5 & 1.63e4 & http://www.archiviolastampa.it/ \\
& 2005 & & &                  3.68e5  & & & & &  \\
 \hline 
Total     &      &      4.23e7 & 3.3e5 & 1.7e6 &1.10e6  &1.12e5& 5.55e5  &6.75e4 &\\ 
 \hline
\end{tabular}
\caption{Size of the online newspaper archives consulted and of the resulting datasets according to the various selection criteria. The totality of the archives has been considered in this work. }
\label{tabellasize}
\end{table}
\normalsize

Only printed editions have been considered and online articles have been excluded for consistency with  pre-internet years.    
The newspapers have freely accessible online archives that cover a major part of their printing time.  Out of the $T\simeq 42$ million articles comprising the four newspaper  online archives, all the articles  which matched the   considered lemmas  ({\it `killed'},  {\it `morti'}, {\it `morta'}  {\it `morto'} {\it `suicide'} ({\it `suicidio'} $\simeq 3.2$ million )  have been extracted.  See Table \ref{tabellasize} for details on the data size of each newspaper. Details of the newspapers and archives are as follow:

\begin{enumerate}	
\item{\it The New York Times} 
  (NYT) was founded in September 1851, with the online archive (https://www.nytimes.com/search/ ) starting in January 1952.
Due to strikes, it was not printed between:  December 9, 1962 - March 31, 1963 (a western edition is present in the archive); September 17, 1965, - October 10, 1965 (an international edition was printed and is present in the archives); August 10, 1978 -  November 5, 1978 (no editions present in the archives).
In 2017, print circulation of the newspaper was 571,500 copies\footnote{New York Times Company form 10-K, 2017.}.
The newspaper began its publications with about 20,000 articles/year  growing gradually  to reach 130,000 articles in 2016. 

\item{\it Il Corriere della Sera} (CDS). 
	The first issue  dates back to March 5th 1876 and is available in  the online archive (http://archivio.corriere.it/). Originally an evening newspaper, it became a morning paper in 1888 and was issued twice a day since 1892 and up two to three times a day in the first part of the XXth Century, but it has been printed as a daily newspaper for several decades. Its offices were bombed on  14/2/1943. Following the liberation of Italy from Nazi occupation, the Committee of National Liberation ({\it Comitato di Liberazione Nazionale}) suspended its publications between   27/4 - 21/5 1945. It resumed publications under the name `Corriere d'Informazione' and, from 1946, as {\it Il Nuovo Corriere della Sera}, with a one-page edition\cite{centoanni,corsera}. 
	CDS passes from $\simeq $7,000 articles/year of  the first years to $\simeq $125,000 articles/year in 2017. The FAQ of the archive reports that it contains  about 2.5 million of pages scanned. In December 2017 it printed 310,275 copies\footnote{ Data on number of printed copies retrieved from http://www.adsnotizie.it  }. 
	
	\item{\it La Repubblica} (REP)
	  begun publications in January 1976, but the online archive \\ 		
		(https://ricerca.repubblica.it/ ) starts on January 4th, 1984, with the first entry of ``morti'' occurring on March 4th, 1984. In December 2017 it printed 274,745 copies.
	REP starts with 24,000/year in 1985 to reach 377,000 articles in 2017.
	
\item{\it La Stampa} (STA).
	The digitalization of this archive (http://www.archiviolastampa.it/)  was performed by the Committee for the Journalistic Information Library ({\it 
 Comitato per la Biblioteca dell'Informazione Giornalistica CBDIG)}.   The archive is released under a Creative Commons license   and covers the period from the first issue, February 9th, 1967, when it was called {\it Gazzetta Piemontese} ({\it Piedmont Gazette}) to December 31st, 2005. Except for a few entries in 1882, all articles up to and included 1909 are referenced in image form with the title {\it  `Notizia'} ({\it `News'}) and no further information on the contents of the article. Several entries after 1909  are stored in this way, thus reducing the overall time range of the dataset. The Fascist government halted   publications of {\it La Stampa} in the month of October 1925  as a warning to all publishers. A few days after resuming publications (November 3rd 1925), Alfredo Frassati, 	 Director of the newspaper for 25 years, resigned to be replaced by directors gradually more aligned with the government. Following the liberation of Italy from Nazi occupation, the Committee of National Liberation ({\it Comitato di Liberazione Nazionale)} suspended its publications between 28 April  - 17 July  1945   (no entries in the database)\cite{stampafascio,giornalismo}. 	 In December 2017 it printed   208,615 copies. 
STA increases from $\simeq $10,000 articles/year to a maximum of 500,000 articles/year in 2001, decreasing at $\simeq $ 200,000 in 2005, the last year present in the archive. 
 
\end{enumerate}

\subsection{Lemmas considered}
This work concerns the study of the occurrence of the following terms: 
\begin{itemize}
\item {\it \bf `Killed'} (in English) and {\it  \bf `morti'} (plural for Italian). 
Note that the literal translation of {\it  killed} in Italian language is {\it  uccisi}, but this word is less frequent than {\it  morti} and usually (but not always) employed in association with murder. {\it  Morti} is more often associated with violent accidental death and thus has a wider correspondence in usage with {\it  killed}.  Articles with the keywords {\it  casualties/vittime},   occur  with a lower frequency than {\it  killed/morti} (10$\%$ to 30$\%$ less). The English term is more strongly  associated with   dead or wounded  during armed conflicts and has peaks during major US wars (Civil, WW1, WW2, Korea, Vietnam etc...). In case of the Italian newspapers, the correlation is lower, since {\it  vittime} is more often  used  for people killed in accidents, natural disasters, and so on. 
\item {\it \bf  `Suicide'} (in English) and {\it \bf  `suicidio'} (in Italian).
\item {\it \bf `morta'}  and {\it \bf  `morto'} (in Italian), respectively feminine and masculine singular form of  `morti'.
\end{itemize}

Some  methods  discussed below can be applied to  any lemma, although the distribution law and last digit analysis is usable in other `quantitative' keywords such as  {\it`dead', `victims', } and the already mentioned {\it  casualties},   will be the subject of future work. 
 
\subsection{Construction of the Data sets}

\begin{figure}[ht]
\begin{center}
\includegraphics[width=1.\textwidth,angle=0]{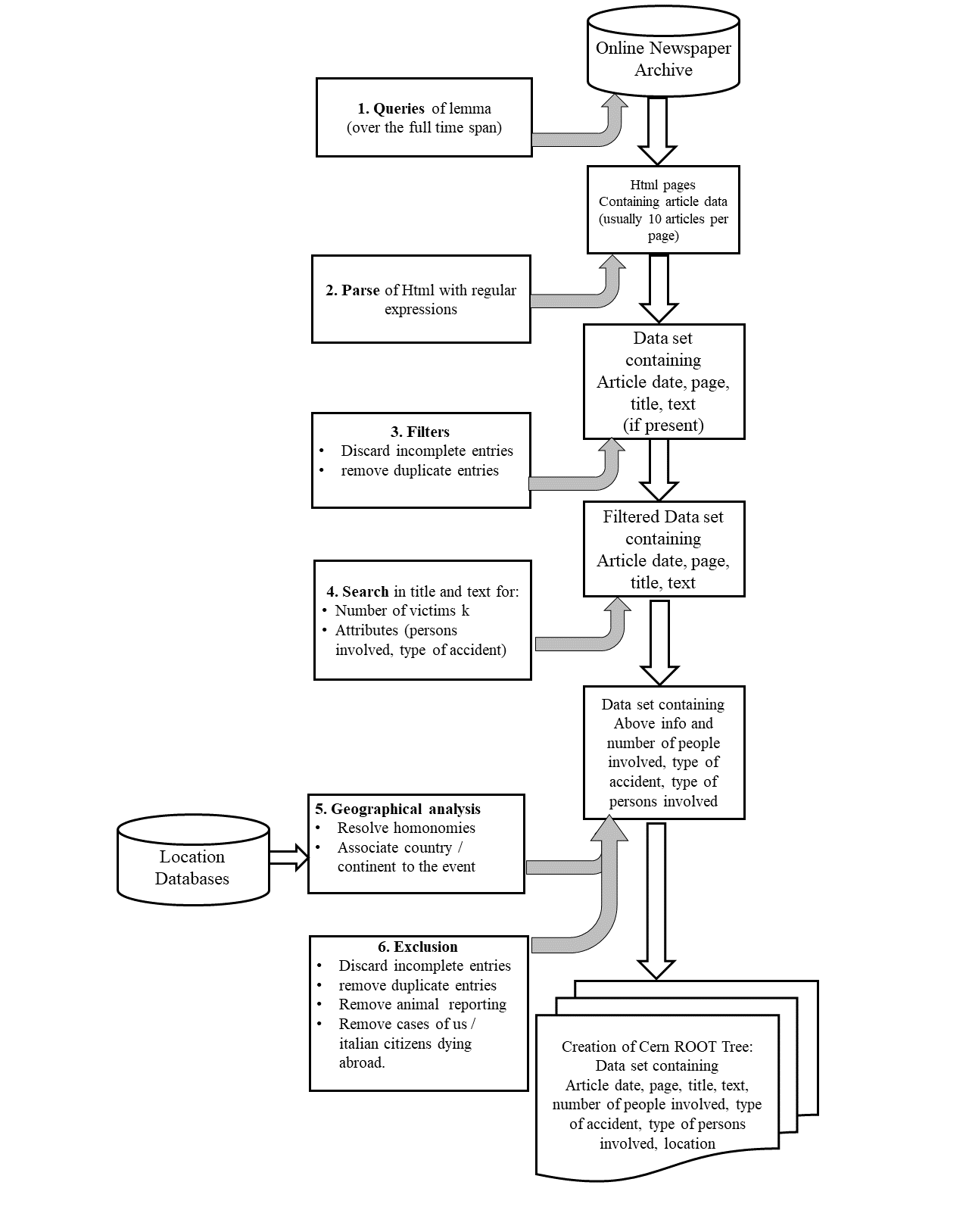}
\caption{Scheme of the processing steps from the query on the  newspaper archive to the creation of the database. Each query of a specific lemma on a newspaper results in a data set which is then subsequently analyzed. }
\label{flow1}
\end{center}  
\end{figure}

{\it 1.  Query and pre-processing. } Each data set has been constructed by querying the web servers of the four newspapers with  Python scripts that emulated manual user input (Figure \ref{flow1}). The input had to be configured for each archive in order to enter the desired  keyword and iterate the requests over the time interval of the newspaper archive. The html pages received in reply to the query were then saved locally. Their overall number depends from the number of articles present in each page: we range from the $\simeq 58,000$ of NYT to the $\simeq 300$ of CDS.  As mentioned, this data acquisition / preprocessing  phase over the newspaper archives   takes 5 to 7 days to complete, depending on the newspaper considered\footnote{We used the selenium (https://www.selenium.dev/) package with python bindings to automatize the interaction with the archives.}. 

We could not  access  the full articles since they are often present only in image form (especially in the years 1850-1910) and the necessary work and resources  would have been almost equivalent to those required for the digitization of the newspaper itself. Furthermore, the work would only have resulted in a marginally higher efficiency in event detection and would not have changed the results.

We define an `event' as  a newspaper  article retrieved upon the query of the keyword killed/morti. In our data sets, each event contains  the newspaper name, the date, and the page (not present in NYT) of the article, the text of the title, and - if present - a part of the text of the article. 

{\it 2. Parse. } The html pages are then  subsequently parsed with Python regular expressions that extract the relevant article information (title, snippet, page number, date, etc...). This  phase takes a few minutes to complete. 
 
{\it 3. Filter. }
Sometimes  the query only returns  a link with no usable text: for instance a title  can be empty  (especially for issues of the  XIX century), incomplete, or does not contain the requested word. These events have been discarded (in the case of STA, this restricts the database to the period 1910-2005 for most purposes).

{\it 4.  Text Analysis.} The events  are then analyzed to find the  number $k$ of people killed. This is searched in numeric, text, and hybrid form. To reduce classification errors, the value of $k$   is searched  close to the keyword   in the forms: {\it `$k$ killed'}, {\it `killed $k$'},  {\it `$k$ attribute killed'}, {\it `killed attribute $k$'}.  
Title and text are also parsed to search for the  type of event (e.g. car, airplane crash, war, illness), and the people involved (e.g. children, women, ethnicity), that will be the subject of a following paper.

{\it 5. Geographical analysis.} Several databases of world places have been used to associate the location event  to its country of origin. Cases of homonymy  (Florence, Paris, Cairo...)  have been resolved assigning the location of the  more famous ones at that time. For instance,  `Cairo' entries in the years 1861-1865 have been assigned to the US since the location appeared often during   the American Civil War.

{\it 6. Exclusions. }
Duplicate entries, that is articles identical in title, text, and date are then removed, but different reports on the same event are considered as separate since it means that they have been considered worthy of more than one article. See Table \ref{tabellasize} for  the size  of the datasets according to the various selections. 

All events reporting deaths of  animals  (fish, herd, cows, etc.), usually associated with high $k$ have been discarded. 

 From foreign events  we  also excluded all  articles where the words `Italian' or `US'/`American'  appear associated with foreign countries, thus removing foreign events where   citizens from the corresponding newspaper-printing country were  involved. This amounted to less than $1\% $ in peacetime.

Once the parsing of the archives is completed, the remaining processing steps in this phase take a few minutes to execute and produce the database / root files for the subsequent analysis. This represents an increase of several orders of magnitute in speed in respect to  any traditional, manual-scanning method which had to be necessarily constrained to a limited amount of time and newspaper issues.

\subsection{Errors }
Statistical errors are due to   fluctuations in the number of events present in a given bin of a given selection. 
Fitting algorithms take into account the errors associated with each bin to calculate the errors of the fitted parameters.  
 
Sources of systematic errors can be due to  OCR (Optical Character Recognition) misidentification. This is more frequent for old issues   where the quality of the scanned  pages is  lower and can result in a lower  efficiency for the first years of the newspapers. 
 
This  can be assumed to be independent of the number, type, or location of people killed $k$ so that the temporal behaviour and distribution laws should not be affected. See the Supplementary Information for a  discussion on systematic errors.

\section{Results and Discussion}

\begin{figure}[ht]
\begin{center}
\includegraphics[width=.95\textheight,angle=90]{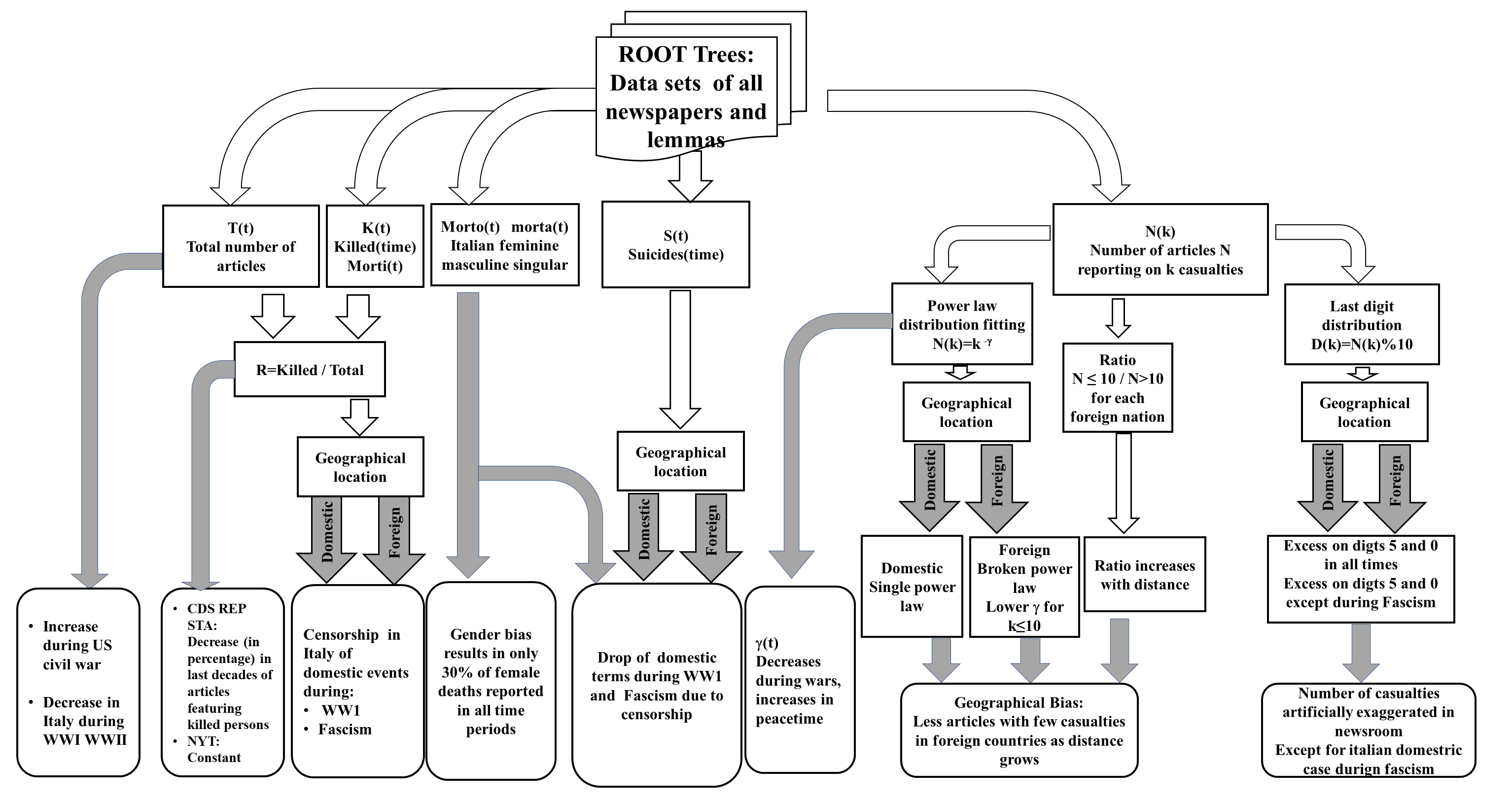}
\caption{Scheme of the various algorithms employed on the databases and the information they provide.  }
\label{flow2x}
\end{center}  
\end{figure}

The   analysis of the data sets created above can yield  information on how the newspapers consider the various  events depending on  geographic location, historical period, or  gender. In this section, we describe the main  algorithms  employed and the  results they provide. The various algorithms, the processing steps, and the key results derived,  are also schematized  in Figure \ref{flow2x}.

 \subsection{Historical  events}   

\begin{figure}[ht]
\begin{center}
\includegraphics[width=.7\textwidth,angle=0]{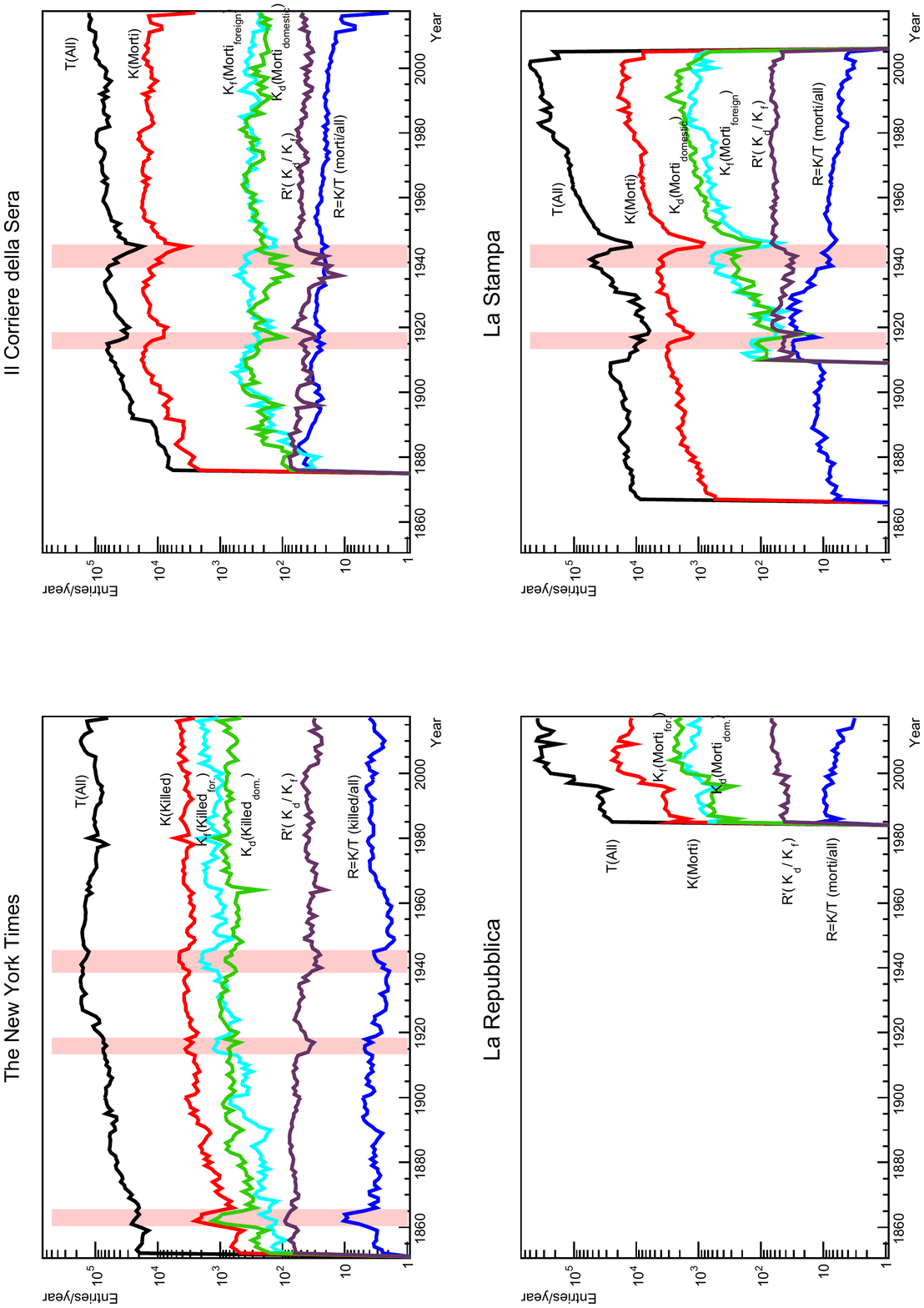}
 \caption{Time profile of datasets retrieved from the  newspaper archives. Each panel shows: a) $T(t)$: Total number of articles present in the archive   (black), b) $K(t)$: number of articles  returned upon query of  the word `killed' or `morti' (red),  c)  $K_{f}$: Number of foreign events (cyan), d) $K_{d}$: Number of domestic events (green), e) $R'= K_{d}/(K_{f})$: Ratio domestic/foreign events   $\times 10^2$ (purple), f) $R=K/T$: (killed/total or morti/total) $\times 10^2$ (blue).  Red bands indicate the time of US Civil War (NYT), WW1 and WW2. It is possible to see the increase of T and K in NYT during Civil War and the corresponding decreases in STA and CDS during the two World Wars. For the same papers it is also evident a gradual decrease of  $K$ and $K'$ during Fascist period (1923-1943) due to censorship (see text).}

\label{morti-norm}
\end{center}  
\end{figure}

  Figure \ref{morti-norm} shows the yearly  total number of articles, $T(t)$. Major historical events can also affect  this value: for instance, it increases in NYT during the US Civil War due to more articles being published and decreases in CDS and STA during the two World Wars  due to due to shortage of materials resulting in fewer pages being published.

In the same Figure the number of articles with  the keyword `killed', $K(t)$, is also shown. From it, we can derive the normalized fraction of  `killed' events in respect to the total:  $R=K/T$ (shown in the same Figure), a value  more affected  by historical events.

For the events with a determined location,   it is possible to separate  domestic ($K_d$) from foreign ($K_f$) occurrences and assess how their relative importance evolved with time (Figure \ref{morti-norm}).  NYT reporting on foreign deaths grows over time to become more frequent than domestic at the onset of  WW1 and permanently from WW2 on. The Italian newspapers divide the reporting between domestic and foreign cases roughly equally, with CDS covering more foreign events in the more recent years and STA and REP the domestic cases.  The other relevant  features are the foreign peaks during the World Wars and the drop in the  domestic deaths between 1923 and 1945 for STA and CDS, due to censorship from the Fascist government (see below).

In Figure  \ref{xxcontinentixx} are shown the relative contributions of the various continents and their evolution over time, showing   a gradual reduction of the coverage of European events and a growing importance of Asia after WW2.    Some major occurrences are:

\subsubsection{\it American  Civil war (12/4/1861-13/5/1865)}

During the American Civil war, the total number of NYT articles  increases   by  $ 18.5\%$ in respect to the interpolated values of $T$ between 1861 and 1865.
 $K$ increases to a maximum  of $K_{1863}/K_{1860}=2.80$ the pre-conflict value (the keyword {\it `casualties'} has an  increase of 
$C_{1863}/C_{1860}=4.23 $).  Conversely, the number of suicides drops to a minimum of $S_{1864}/K_{1860}= 0.36 $.

\subsubsection{\it World War 1 (28/7/1914-11/11/1918, Italy from 23/5/1915, US from 6/4/1917)}  

Censorship was very strong in all countries involved in both World Wars, with the removal of  all information who could be beneficial to the enemy: letters, reporting of battles and defeats, casualties etc.\cite{book206526}.

 In NYT the value of  $R=K/T$   increases  from 
$R_{1913}=0.361 \pm 0.001 $   to $R_{1914}=0.457\pm 0.001$. The domestic event ratio $R'=K_{dom}/K_{total}$ drops from $R_{1913}^{'\: NYT}=0.56\pm 0.02 $ to $R_{1914}^{'\: NYT}=0.39\pm 0.01 $,  reaching a minimum of  $R_{1917}^{'\:  NYT}=0.32\pm 0.01$, when the US declared war to Germany.

 In Italy, the  shortage of resources resulted in a reduction of the number of pages of CDS and STA from 8 (two double sheets) to 4 (one double sheet of paper).  Consequently, $T$ decreased to $T^{CDS}_{1918}/T^{CDS}_{1915}=0.48$ and $T^{STA}_{1918}/T^{STA}_{1915}=0.67$. 
The effect of censorship is evident in STA: its value of $R$ drops from $R_{1915}=0.32\pm 0.02$  to a minimum of  $R_{1917}=0.13\pm 0.01$ (for CDS is more constant). In both newspapers, there is an even  higher drop in domestic events (not necessarily  only due to censorship but also to a lack of interest): from $R_{1914}^{'\:  STA}=K_{dom}/K_{total}= 0.43\pm 0.07 $   to $R_{1917}^{'\: STA}=0.26\pm 0.09$. For CDS, R drops from $R_{1914}^{'\: CDS}=0.51 \pm 0.06$ to $R_{1917}^{'\: CDS}=0.29 \pm 0.04$.

During wars the rate of suicides is known to decrease: this phenomenon is usually explained by the higher sense of purpose during the bellic effort\cite{doi:10.1111/j.1943-278X.1971.tb00598.x,somogyi,10.2307/23636204} and is found to occur both  when one's country or  other countries are at war. However, the decrease of suicides reporting by newspapers  is more prompt and intense (dropping to 1/3 of the pre-war value) than  that recorded by statistics.  Since in both countries this occurred in 1914, before either country was at war we can conclude that this was not directly related to censorship. 
 In Italy, the number of articles on suicides passes    from $S^{STA}_{1913}/T^{STA}_{1913}=0.043\pm 0.002 $    to $S^{STA}_{1914}/T^{STA}_{1914}=0.032\pm 0.002 $  and reaches a minimum of $S^{STA}_{1917}/T^{STA}_{1917}=0.012\pm 0.001$   before returning to $S^{STA}_{1919}/T^{STA}_{1919}=0.025 \pm 0.002$.
A similar behaviour is found for  CDS (and NYT):  from $S^{CDS}_{1913}=1.34\pm 0.06$ ($S^{NYT}_{1913}=1.18\pm 0.04$)  to $S^{CDS}_{1914}=0.73\pm 0.04$ ($S^{NYT}_{1914}=0.92\pm 0.04$ ) at the beginning of WW1 to a minimum  $S^{CDS}_{1916}=0.45\pm 0.03$ ($S^{NYT}_{1916}=0.39 \pm 0.02$).
The drop before the  US or Italy entered the war suggests to attribute  the lack of suicide reporting to a reduced interest by the editorial rooms rather than to censorship.

\subsubsection{\it Fascist government in Italy (31/10/1922 - 25/7/1943)}

Different is the case during the Fascist government in Italy. The Italian government of the time  exercised a strong censorship on  printed press and radio. On 14/7/1924 a {\it circolare} (note)  from the then Minister of Interior Federzoni allows the sequestering of copies of newspapers to 'prevent stirring up  public opinion'. On 31/12/1924 all newspapers are sequestered and the directors replaced with ones affiliated with the regime. 
In October 1926, several daily newspapers  were closed until the end of WW2. Among them  {\it L'Unit\`{a}},   {\it L'Avanti!} and {\it L'Ora}\cite{corsera,centoanni}.

Government censorship aimed  to present an efficient state and thus had to remove all negative news.  Censorship involved all media of the time: radio, theater, movies, books, and newspapers. Authors, especially those of Hebraic origin but also those who were against the regime for political reasons, fell in disfavour. This `targeted' censorship was similar to what occurred in Germany and reported in\cite{ScienceMichel176}, where prominent individuals   were  mentioned to a greater or lesser  extent according to their race or standing in respect to the Nazi government.   

On a wider scale, government guidelines\cite{stampafascio, giornalismo} to newspapers required that  crime reporting  had to be compressed in a few lines, and suicides had  to be  ignored, with the result that articles involving domestic deaths and accidents almost disappeared from newspapers.   
With the datasets of CDS and STA it is possible to quantify  the overall effect of Fascist censorship in the reporting of violent deaths \cite{defelice,aquarone}.

As a result, even though the value of  $R=K/T$ remains more or less constant, in 1923 $R'$  drops  from $R^{'\: CDS}_{1922}=0.56 \pm 0.03 $  ($R^{'\:  STA}_{1922}= 0.64\pm 0.05 $)   to $R^{'\: CDS}_{1923}=0.44\pm 0.02 $ ($R^{'\:  STA}_{1922}=0.46\pm 0.05 $)  reaching   a minimum of $R^{' CDS}_{1936}=0.119\pm 0.005 $   ($R^{' STA}_{1937}=0.31\pm 0.01 $) (Figure \ref{morti-norm}). These values come back to pre-dictatorship values of  $R^{' CDS}_{1946}=0.60 \pm 0.03 $ ($R^{' STA}_{1946}=0.68\pm 0.06 $) immediately  after the war when both newspapers had their publications halted  and their directors were replaced between April and May (July for STA)  1945. 

 We estimated the amount of censorship for articles with {\it morti} interpolating  the value of $R'$ between  1922 and  1946: between  1923 and 1943  there were 2,800$\pm 200$ domestic articles with at least two casualties  missing for CDS and 2,900$\pm 300$ for STA. In both cases, they amount to $75\%$ of the articles featuring domestic deaths printed in the same period. 

A similar analysis on the $k=1$  events ({\it morto, morta}, see Figure \ref{MORTOMORTAMORTI})  yields about $22,000\pm 1300$ articles  censored by STA ($60\%$ of those printed in the period considered) and  $29,000\pm 500$ 	 by CDS ($20\%$ of those printed in the period considered).  

\begin{figure}[!ht]
\begin{center}
\includegraphics[width=0.9\textwidth]{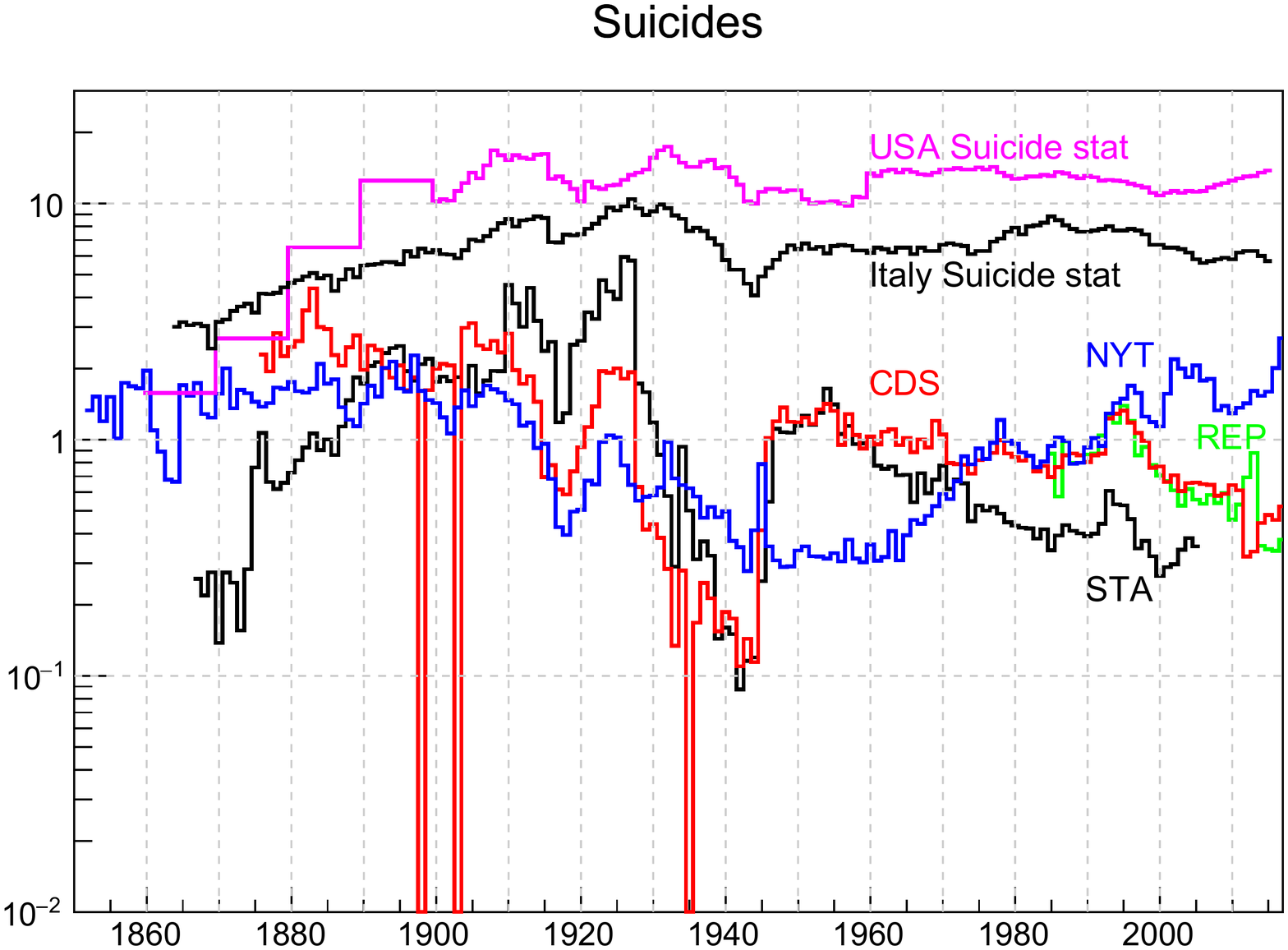}
 \caption{Percentage of articles containing the keyword  {\it 'suicide/suicidio'} for the four newspapers and the rate of suicides (cases per 100,000 population) vs time\cite{somogyi,vital,oecd}. Note the sharp drops during the American Civil War (for NYT only), WW1 and WW2, both in the articles and in the rate. The drop during Fascist government is also visible. In 1898, 1902 and 1935 there are no articles retrieved with the keyword. }
\label{suicidio}
 \end{center}
\end{figure}

In the case of  suicides   (another forbidden topic during the regime\cite{corsera}), Figure \ref{suicidio} gives $S_{missing}=17,000\pm 600 $   ($S_{missing}/S_{present}=2.8$) for CDS and 4,100$\pm300$ ($S_{missing}/S_{present}=1.1$) for STA,  in contrast with  the growing rate of  suicides at the time\cite{somogyi, suicidio1971}.           

Overall we estimate that CDS censored $41,800\pm 1000$  articles and STA $36,000\pm 1,900$ during this period, with an average of $1990\pm 50/year$ for CDS and $1700\pm 90/year$ for STA.

\subsubsection{\it  World War 2 (1/9/1939 - 2/9/1945, Italy from 10/6/1940 as part of the Axis and from 8/9/1943 as part of the Allies, USA from 7/12/1941)}

Censorship in US  during WW2 relied mostly upon the self-censorship of news outlets: forbidden topics included weather and crop reports, correspondence, travel schedules and naturally troop movements\cite{book1401476}. The Office of Censorship released on January 15, 1942, the Code of Wartime
Practices for American Broadcasters and the Code of Wartime Practices for the American Press.  The publication of any pictures depicting US soldiers  killed in combat was forbidden until September 1943, when the capitulation of Italy might have induced in the general public the idea that the war was to end soon\cite{book206526}. 
  
 The total number of articles of NYT  remains   essentially unchanged, with $R$ increasing only in 1940, from $R_{1938}=2.02\pm 0.04\%$ to  $R_{1940}=2.50\pm 0.04\%$. This is due to an increased amount of reporting of casualties prior to the entrance in the conflict. This is confirmed by the value of $R'$, from $R'_{1938}=K_{foreign}/K_{domestic}=0.63\pm 0.01$ to a maximum of $R'_{1940}=K_{foreign}/K_{domestic}=0.76\pm 0.01$, again before Pearl Harbor. Afterward, since US soil was not    attacked, the amount of reporting  of   domestic events  was not affected so we can conclude it was not affected by censorship. 

The effect of WW2 in Italy is mostly evident in the sharp drop in  $T$ and $K$ in the later years of the war. This was both due to paper shortage and the bombings on Milano and Torino, where the newspapers were printed. 
After the armistice with the Allied forces (8/9/1943),   Italy was  split between the South, controlled by the Allies and  {\it La Repubblica di Sal\`{o}} in the North. After a short period free from Fascist government, CDS and STA are then aligned to the Nazi-controlled  government of the North so the definition of `domestic' and `foreign' becomes fuzzier and the ratio $R'$ increases.

\subsubsection{Recent decades}
If we consider the reporting of deadly accidents in the last decades we see that it is constant or decreasing: in the NYT, after doubling between 1960 and 1969, $R^{NYT}$  remains - with fluctuations - constant up to 2017. 

 The Ratio $R$ for STA  decreases gradually after 1954, with $dR/dt_{STA\:  1954-2005}=(-0.1\pm 0.02) \% /year$.  

Also CDS exhibits a similar but stronger decrease:  $dR/dt_{CDS\: 1954-2005}=(-0.3\pm 0.02) \%/year$.

For REP, the amount of articles reporting violent death   $K$  increases with time but  its percentage after 1995   decreases by    $dR/dt_{1995-2005}=(-0.2  \pm 0.04 )\%/year$.  
We note that this  trend of decreasing coverage given to violent events in all three Italian newspapers is opposite to the growth of the  perceived threat of violence in Italy\cite{ISTAT}, so this phenomenon cannot  be ascribed directly to press coverage.

\subsection{Gender Bias}

In Italian newspapers, where the language allows to distinguish   gender, we have also queried the databases for the term {\it `morto'/`morta'} ($M$), respectively the singular ($k=1$) masculine, and feminine form in the Italian language. {\it `Morte'}, the feminine plural could not be used since it also means `death' in Italian and it would be difficult to semantically separate the two meanings. Besides, if both males and females are involved the term {\it 'morti'} is used in Italian language, making the lemma {\it 'morte'}  of little use.

\begin{figure}[!ht]
\begin{center}
\includegraphics[width=0.49\textwidth,page=1]{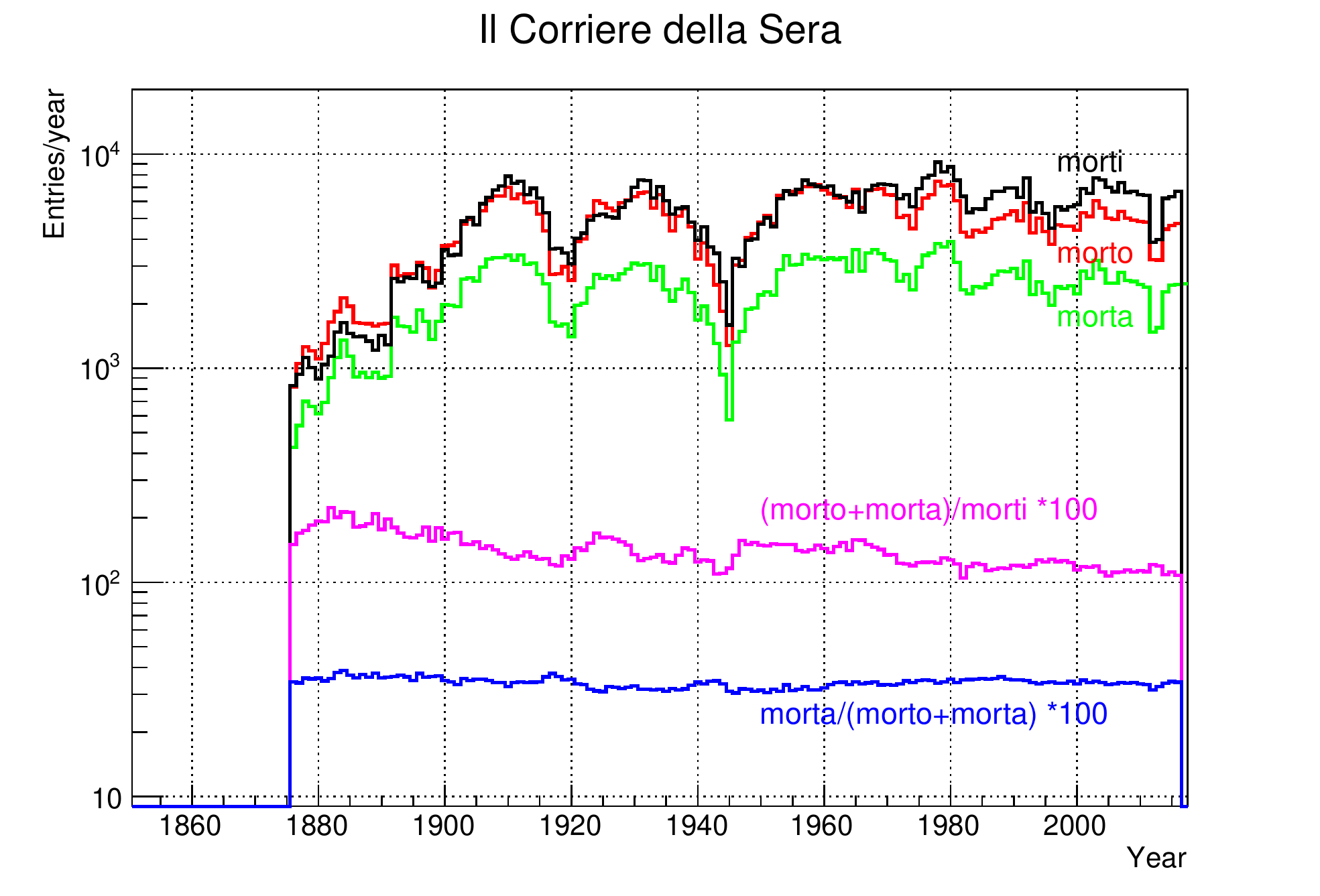}
 \includegraphics[width=0.49\textwidth,page=3]{./fig5.pdf}
 \includegraphics[width=0.49\textwidth,page=2]{./fig5.pdf}
  \caption{ Number of occurrences for singular ({\it morto, morta}) vs plural ({\it morti}). Magenta is the ratio $(k=1)/(k\geq 2)$, ({\it morto+morta)/morti}. The blue ratio refers to the percentage of female deaths, {\it morta/(morto+morta)}. Both ratios are  $\times 10^2$).  
 The percentage of $k=1$ events halves for CDS over time and is roughly  constant   for STA after $\simeq 1890$, with drops during WW1 and Fascism. The reporting of female deaths is constant to $\simeq 30\%$ for CDS and   STA, and doubles from $20\%$ to $40\% $ for REP.  }
 \label{MORTOMORTAMORTI}
\end{center}
\end{figure}

From Figure \ref{MORTOMORTAMORTI}  we see that the amount of reporting of  female deaths ({\it morta}) is  only $30 \%$ of  all $k=1$ deaths. 
This has to be compared with the fatal accident standardized death rate (in 2005) in Italy of 36.1 (male) and 19.2 (female)  per 100,000 deaths \cite{EUdeath}, and the probability for a 15-year old individual to  die within 45 years, before reaching the age of 60 (45q15)\cite{Rajaratnam2010} in 2010 of $7.9\%$ for men  and  $4.1\%$ for women. 

Therefore, even accounting for the fact that male violent deaths are more frequent than female ones\footnote{For instance, in US the number of fatal occupational injuries for US in 2016 involved women in $7\%$ and men in $93\%$ of cases, for an almost equal  number of worked hours ($43\%$ for women and $57\%$ for men)\cite{USWorkCensus}.} and that these events would be more likely to be reported in the news, this still hints to some amount of gender bias in reporting. The female/male ratio of $\simeq 30\%$ is present in all newspapers and roughly constant over the years, with only REP showing  an increase of reporting of female deaths of about $3\%/year $, from $22.6\%$ in 1985 to $37.65\%$ of 2007.

\subsection{Scaling laws}

The analysis of  keywords associated with the number of persons involved allows  
 to  build  the distribution function of the number of articles, $N_k$, reporting $k$  people killed. As shown in Figure \ref{tuttimortinogeolocALL}, the distributions for all four newspapers considered  can be described by a single power-law:
\begin{equation}
N_k=N(k)=A\cdot k^{-\gamma}
\end{equation}
in the range $2\leq k \leq 10^6$. The sharp peaks in $N_k$ for values of $k$ that are multiples of 10, 100, 1000... are due to the  rounding in excess to the nearest multiple of a power of 10 of the number of people reported (see below).

The Minuit\cite{James:1994vla} package (now in its second release, Minuit2) has been used  to perform the fits.  

Fitting of the power laws has been analogous to  the methods that we used in the fitting of cosmic ray spectra of the PAMELA space-borne detector\cite{Adriani2014323}, \cite{2011Sci...332...69A}, \cite{2009Natur.458..607A} (see also ext. data therein). 
Tests with varying bin sizes have shown no significant change in the fitting results and values of $\gamma $.
 
See Section \ref{mathappendix} for the mathematical details; a discussion of the fitting methods and error  systematic can be found in the Supplementary Information.

Power-law statistics has been found to  describe  the distribution of various natural phenomena, e.g.  earthquakes\cite{doi:10.1029/JB094iB11p15635},   forest fires\cite{Malamud1840}, \cite{2007CNSNS..12.1326T},   the   cluster size of tropical trees\cite{Condit1414}, 
and is usually thought to arise through positive growth feedback
\cite{Scanlon2007}, \cite{2005GeoJI.163..433Y}.
 
Galactic cosmic ray spectra also follow a power law, as a result of the statistical process of acceleration. Changes in the spectral indexes show the presence of additional sources or production phenomena\cite{2011Sci...332...69A}, \cite{2009Natur.458..607A}.

Power law distributions are also encountered in many human-related  activities\cite{powerlawdoi:10.1137/070710111}, from language distribution (Zipf law\cite{doi:10.1080/00107510500052444},
 \cite{10.1371/journal.pone.0147073}), the  number of casualties in wars\cite{doi:10.1080/01621459.1948.10483278}, 
\cite{cederman_2003}, 
and  ethnic violence \cite{Lim1540}.
Also in these cases, they have been shown to arise through a `winner takes all' type of a competitive network where a few elements grow to acquire a very large size \cite{Barabasi2005},
 \cite{Bohorquez2009}.

In newspapers, the distribution   $N_k$ can be explained as the result of two main phenomena:
\begin{itemize}
\item  The convolution of various violent events and accidents. Road, train, air accidents, natural disasters and catastrophes have each  their  frequency and  probability distribution, usually 	unknown but decreasing as $k$ increases.   
 
Articles with $k \simeq >  1000 $ often do not describe a specific accident, but rather  summarize  global phenomena such as   war, illness deaths (cancer, heart attack...)  or automobile accidents per year, etc... . 

\item The selection by the newsroom. The publishing criteria can change with time, location or censoring:  events with higher $k$ will have a higher probability of being selected for their importance. Conversely, foreign events, occurring in countries physically or  socially far from the country where the newspaper is printed, will be more likely to be ignored, especially for low 
$k$.  
\end{itemize}

Both phenomena can be approximated by a power law, with probability $P\propto k^{-\gamma_1}$ for an event to involve $k$ casualties, and a  probability (or efficiency)  $\epsilon \propto k^{+\gamma_2}$ for the event to be picked by the newsroom. 
The  overall probability is then  $P_{tot}=\epsilon \cdot P \propto k^{-\gamma_1+\gamma_2}$ with $\gamma=\gamma_1+\gamma_2$.

The values of $\gamma$ of the four newspapers 
($\gamma_{NYT} =  1.44\pm 0.06$,  $\gamma_{CDS} = 1.61\pm 0.01$,   $\gamma_{REP} =1.42\pm 0.07 $,  $\gamma_{STA} = 1.52\pm 0.09$, Table \ref{tabellagamma})
show that the editorial processes are similar in Italy/US and converge to a narrow range of values, although the differences among them reflect the different emphasis to high/low $k$ events in the  newspapers.
A `steep' spectrum, with a high $\gamma$,  results in a higher number of articles with low $k$ and vice-versa. Defining   $W=N_H/N_L=N_{k> 10}/N_{2\leq k\leq 10}$ we see (Table \ref{tabellagamma}) that it ranges between $W_{CDS}=0.57$ (more articles with low k) and $W_{REP}=1.00$ (fewer articles with low k)   with $W_{STA}=0.72$ and $W_{NYT}=0.93$  having intermediate values (see Section \ref{mathappendix} and Figure \ref{theor_ratio} for a calculation of the value of $W$ as a function of $\gamma$).

\begin{figure}[!ht]
 \includegraphics[width=1.\textwidth,page=1]{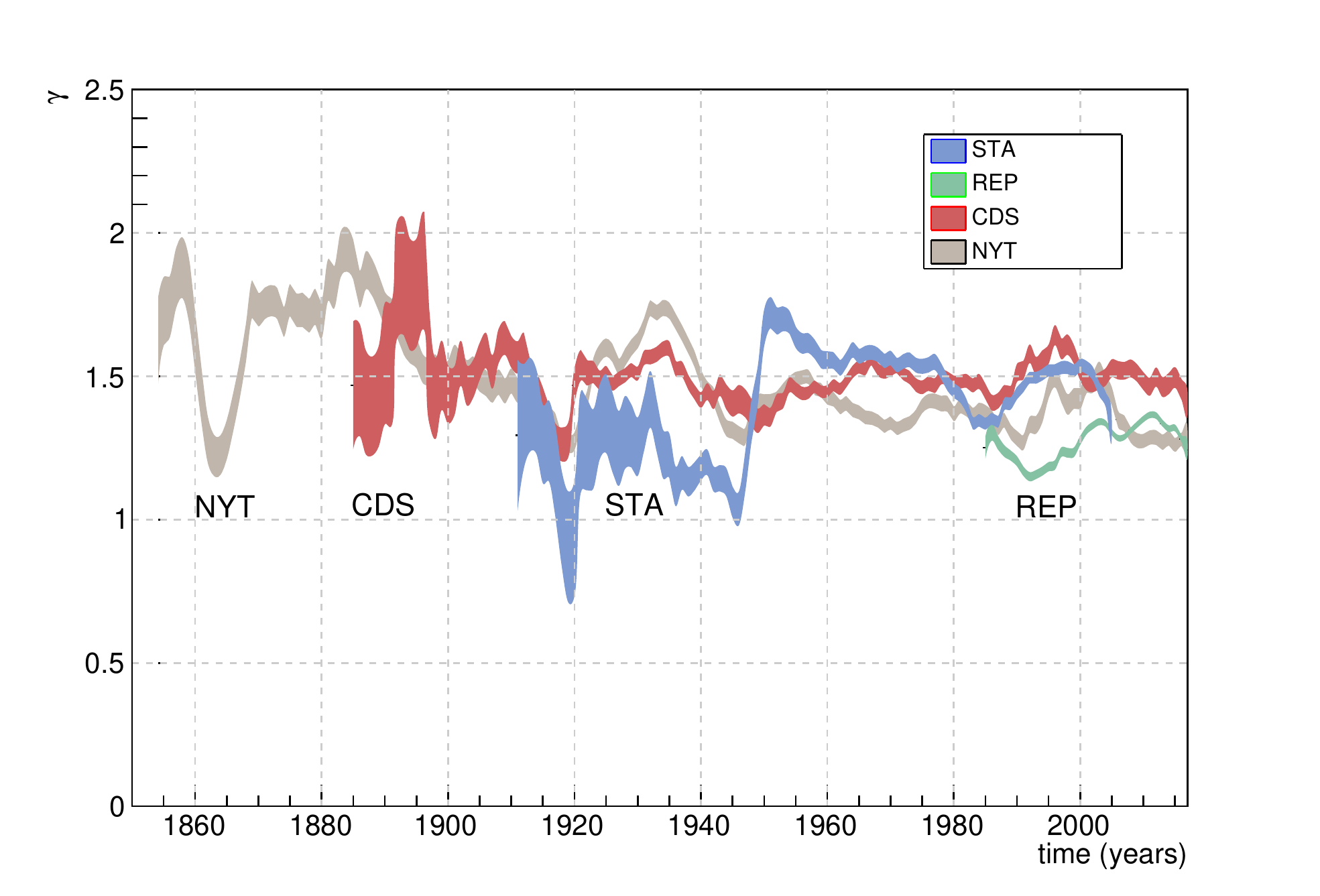}
 \caption{Spectral index $\gamma $ as a function of time (running average on current and 4 preceding years). $\gamma$ decreases during conflicts such as the Civil and the World Wars due to the high $k$ events. The bands indicate the error  from the fit.
   }
\label{gammavsannomm5}
\end{figure}

The trend of the  spectral index can be used to estimate the state of belligerence reported by the newspapers: the running average (current and four preceding years) of $\gamma(t)$ (Figure \ref{gammavsannomm5})  
decreases in  wartime due to the higher abundance of high  $k$ events results in  a flatter distribution.  Vice-versa  in peacetime, when the distributions are dominated by low $k$ events, $\gamma $ increases due to a steeper distribution.  Thus, local minima in $\gamma (t)$  are present in NYT during  the US Civil War ($\gamma = 1.21 $),
the two World Wars ($\gamma_{WW1} =1.26$, $\gamma_{WW2} =1.30$) and the Vietnam War ($\gamma =1.31$).  Also, CDS and STA show similar local minima during WW1 ($\gamma^{CDS}\simeq 1.2$ and $\gamma^{STA}=0.8$) and WW2 ($\gamma^{STA}=1.76$, $\gamma^{CDS}=1.91$), followed by a  sharp increase in STA (and more gradual in CDS) after 1945. 
It is also interesting to note that  - notwithstanding the differences in $\gamma$ -  the trends of the spectral indexes of the newspapers are in  good agreement among each other. This suggests that they all  tend to react similarly   to  the conflicts occurring  in  the world (and vice-versa  a discrepancy in the trend would indicate the presence of censorship).

	\begin{table}[t]
\begin{center}
\begin{tabular}{c c c c c c c}
 \hline
Newspaper 	& Loc. & $\gamma$           &  	$\chi^2 /NDF  $  	& $N_{H}/N_{L}$   \\
 	\hline
	NYT  & All &1.44$\pm$   0.01  &9  &0.93 $\pm $0.01 \\
CDS  & All &1.61$\pm$   0.01  &7  &0.57 $\pm $0.01 \\ 
REP  & All &1.42e$\pm$   0.01 &16 &1.00 $\pm $0.01 \\
STA  & All & 1.52$\pm$   0.01 &9  &0.72 $\pm $0.01 \\
  \hline
\end{tabular}
\caption{Spectral index $\gamma$  for a single power-law fit ($2\leq k \leq 10^6$) on the full  datasets of the four newspapers. }
\label{tabellagamma}
\end{center}
\end{table}

	\begin{table}[t]
\begin{center}
\begin{tabular}{c c c c c c c c}
\hline 
Newspaper 	& Loc. & $\gamma_L$     & $\gamma_H$        &    $\Delta\gamma $        & 	$\chi^2 /NDF $ & M  	    \\
            &      &                 &                  &   $\gamma_L-\gamma_H$       &           &  (Excess/Defect ) \\ 
 	\hline
NYT  & Dom. &1.37e+00$\pm$   3e-03&1.78  $\pm$   1e-02  & -0.417$\pm$   1e-02  &    3e+01  &        4$\%$ \\
NYT  & For. &9.90e-01$\pm$   3e-03&1.66   $\pm$   5e-03  &-0.672$\pm$   5e-03  &    6e+01  &      -98$\%$ \\
CDS  & Dom. &2.08e+00$\pm$   3e-03&1.81   $\pm$   2e-02  &0.277$\pm$   2e-02  &    2e+01  &       71$\% $\\ 
CDS  & For. &5.02e-01$\pm$   3e-03&1.59   $\pm$   5e-03  &-1.09$\pm$   6e-03  &    1e+02  &     -122$\% $\\ 
REP & Dom. &1.69e+00$\pm$   3e-03&1.54   $\pm$   6e-03  &0.143$\pm$   7e-03  &    2e+02  &       16$\% $\\
REP & For. &9.66e-01$\pm$   3e-03&1.48    $\pm$   4e-03  &-0.511$\pm$   5e-03  &    1e+02  &      -91$\% $\\
STA & Dom. & 1.95e+00$\pm$   2e-03&1.72  $\pm$   1e-02  &0.2311$\pm$   1e-02  &    4e+01  &       47$\% $\\
STA & For. & 9.23e-01$\pm$   3e-03&1.54   $\pm$   5e-03  &-0.615$\pm$   6e-03  &    9e+01  &      -94$\% $\\
 \hline
\end{tabular}
\caption{Spectral index for the four newspapers separating domestic and foreign datasets. $\gamma_L=\gamma(2\leq k \leq 10)$, $\gamma_H=\gamma(10\leq k \leq 10^6)$. A negative $\Delta\gamma=\gamma_L-\gamma_H$ implies a lack of events for $2\leq k\leq 10$ events, and vice-versa. The higher the absolute value the higher the excess or defect of events. $M$  is  the excess (+)/defect (-) of articles with respect to what expected from a pure power-law (see text).}
\label{tabellagammadeltagamma}
\end{center}
\end{table}

\subsection{Geographical bias}

\begin{figure}[!ht]
\begin{center}
\vspace{-1.5cm}
\includegraphics[width=0.81\textwidth,page=1]{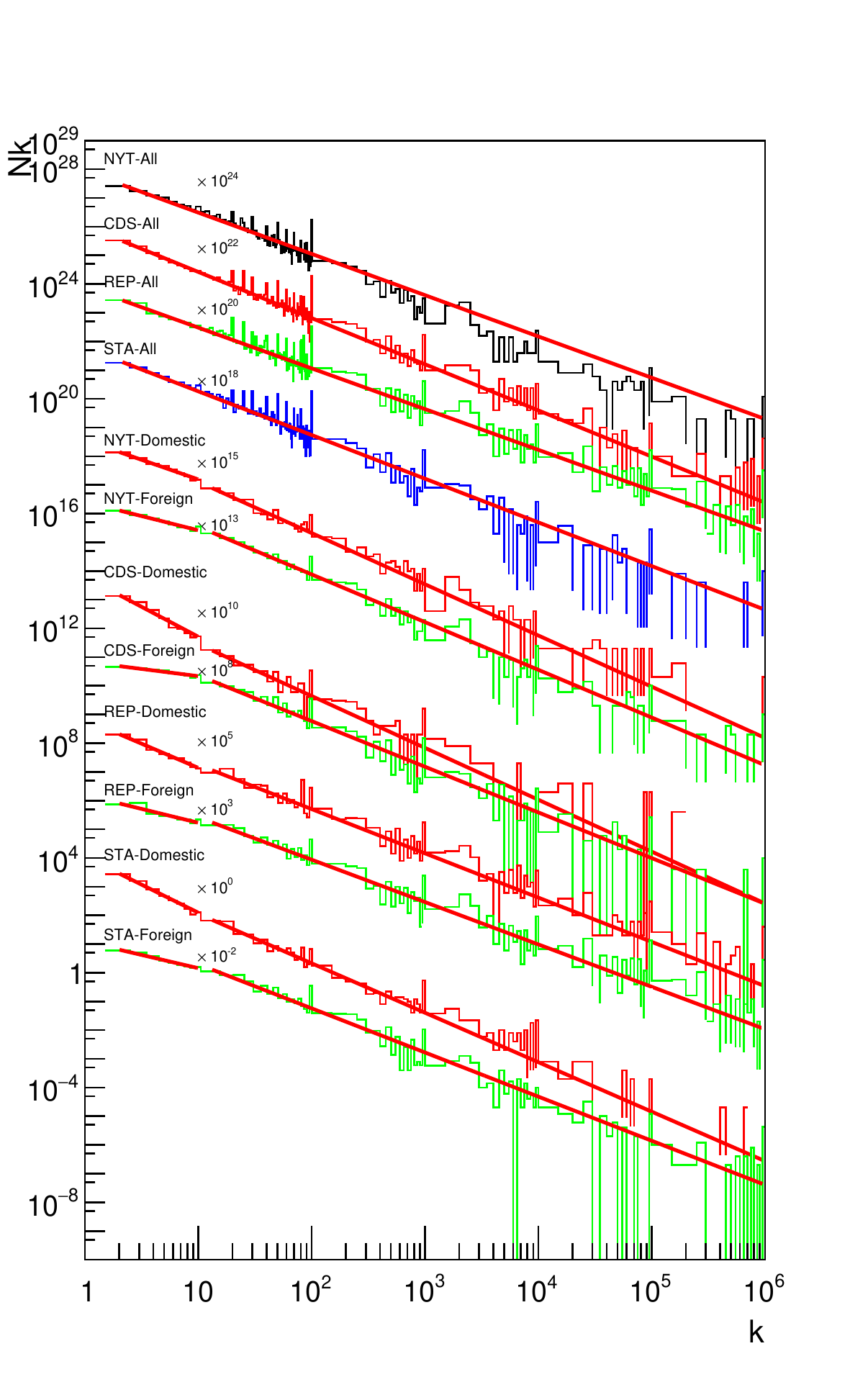}
 \vspace{-1.2cm}
\caption{ Number of articles $N_k$ as a function of persons killed $k$ mentioned in each article. The full distributions have been fitted with a single power law (values in Table \ref{tabellagamma}), the domestic/foreign ones with two power laws $\gamma_L$ ($2\leq k \leq 10$) and $\gamma_H$ ($k>10$). For all foreign events there is a  break in the spectral index $\gamma_L<\gamma_H$ due to missing events not being reported (from $91\%$ for REP to $122\%$ for CDS). In domestic events $\gamma_L>\gamma_H$ for Italian papers (over-reporting of low k events, from $47\%$ in STA to $71\%$ for CDS). In NYT whereas the decrease in $\gamma_L$ for domestic events is lower (over-reporting of $4\%$) than for foreign ones (under-reporting of $98\%$), hinting to a higher degree of internationalization of this publication (Table \ref{tabellagammadeltagamma}). }
\label{tuttimortinogeolocALL}
\end{center}
\end{figure}

\begin{figure}[!ht]
\begin{center}
\includegraphics[width=1.\textwidth,page=1]{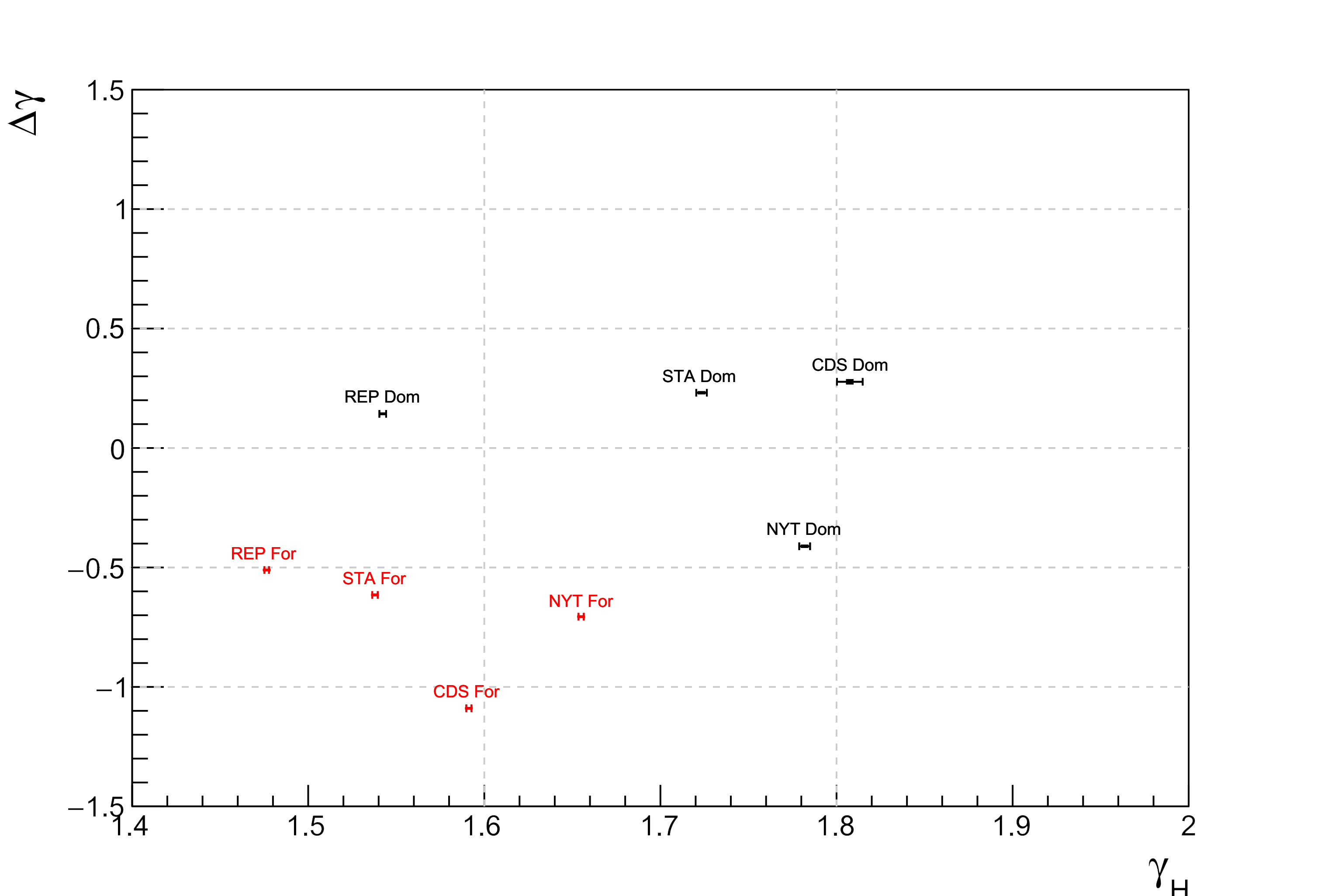}
 \caption{  $\Delta \gamma=\gamma_L-\gamma_H$ vs $\gamma_H$  obtained from the fit of power-law for the distributions of domestic and foreign events. $\Delta \gamma\simeq 0$ means a pure power law over the  whole distribution whereas a negative (positive) $\Delta \gamma$  implies  a lack (excess) of articles with a small ($2\leq  k\leq 10$) number of deaths.   An higher (lower) value of  $\gamma_H$ implies more emphasis on low (high) values of k. In all newspapers, foreign distributions have a lower $\Delta \gamma$ than domestic ones and     a lower value of $\gamma_H$, showing that high $k$ events have an higher importance over the low $k$ ones. NYT has the smallest differences in $\Delta \gamma$, a sign of greater uniformity  of treatment between domestic and foreign events.      }
\label{NDgG}
\end{center}
\end{figure}

In our case, using the articles where a location could be determined, we built distribution laws for domestic and foreign events.  In Figure \ref{tuttimortinogeolocALL} it is possible to see that the latter have a strong spectral break for $2\leq k\leq 10$ (absent or less prominent in the former). Fitting the distribution function with two power laws $\gamma_H=\gamma(k>10)$ and $\gamma_L=\gamma(2\leq k\leq 10)$ we see that $\Delta \gamma=\gamma_L-\gamma_H$ is always negative for foreign events, ranging from  $\Delta \gamma_{REP}=-0.511\pm 0.005$ to $\Delta \gamma_{CDS}=-1.09\pm 0.01$. For Italian newspapers featuring domestic events,  the distributions are closer to a pure power law (highest $\Delta \gamma=+0.28\pm 0.02$ for CDS). In all  cases $\Delta \gamma$ is positive, implying that  a higher emphasis is given to low $k$ events.
  NYT has a smaller discrepancy between foreign ($\Delta \gamma=-0.672\pm 0.003$)  and domestic events ($\Delta \gamma=-0.42\pm 0.01$). A plot of the value of $\gamma $ $vs$ $\Delta\gamma $ (Figure \ref{NDgG}) shows that the foreign and domestic  categories  are clearly separated  in all four newspapers (Table \ref{tabellagammadeltagamma}). This suggests a difference in  the editorial behaviour  due to a lack of press coverage of  accidents involving a small number of persons in foreign  countries,   considered   to be not interesting enough to be reported in the press.

An estimation of the under- or over-reporting of low $k$ events can be provided by the extrapolation of   $\gamma_{H}$ to $2\leq k \leq 10$:
\begin{equation}
M=\frac{N_L - \int_{2}^{10}\alpha_{H} k^{-\gamma_{H}} dk  }{N_L}
\end{equation}

with $N_L= \Sigma_{2}^{10}N_k$ and $\alpha_{H}$ coming from the power law fit of $k\geq 10$. Thus, M is the  fraction of events with $2\leq  k\leq 10$ missing ($M<0$) or exceeding ($M>0$)   the value expected from a pure power law distribution ($M=0$). 
All four newspapers have $M_{for}\simeq -100\% $ in the foreign case, meaning that the editorial room decides to print  only one event out of two if it involves ten or fewer casualties in a foreign country. 
We also note that Italian newspapers tend to print more news of domestic events (from  $M_{dom}=16\% $ of REP  to $M_{dom}= 71\%$ of CDS) than what expected from a pure power-law distribution. This can be attributed in part to a large domestic and local news   section. Overall 
CDS has the largest difference in dealing with foreign and domestic events ($M_{for}^{CDS}=-122\%$ - $M_{dom}^{CDS}=+71\%$ ) and the NYT the smallest ($M_{for}^{NYT}=-98\%$ - $M_{dom}^{NYT}=+4\%$ ). See Section \ref{mathappendix} and Figure \ref{missing} for a calculation of the value of $M$ as a function of $\delta \gamma$ and $\gamma_H$.

In many nations, there are too few events reported to perform an acceptable fit with a power law, 
 therefore - for countries having  at least 30 entries in a given newspaper - we used  the ratio the $W=N_{2\leq k\leq 10}/N_{k>10}$ as an indicator of the intrinsic importance assigned to a given country.
 
\begin{figure}[!ht]
\begin{center}
 \includegraphics[width=.8\textwidth]{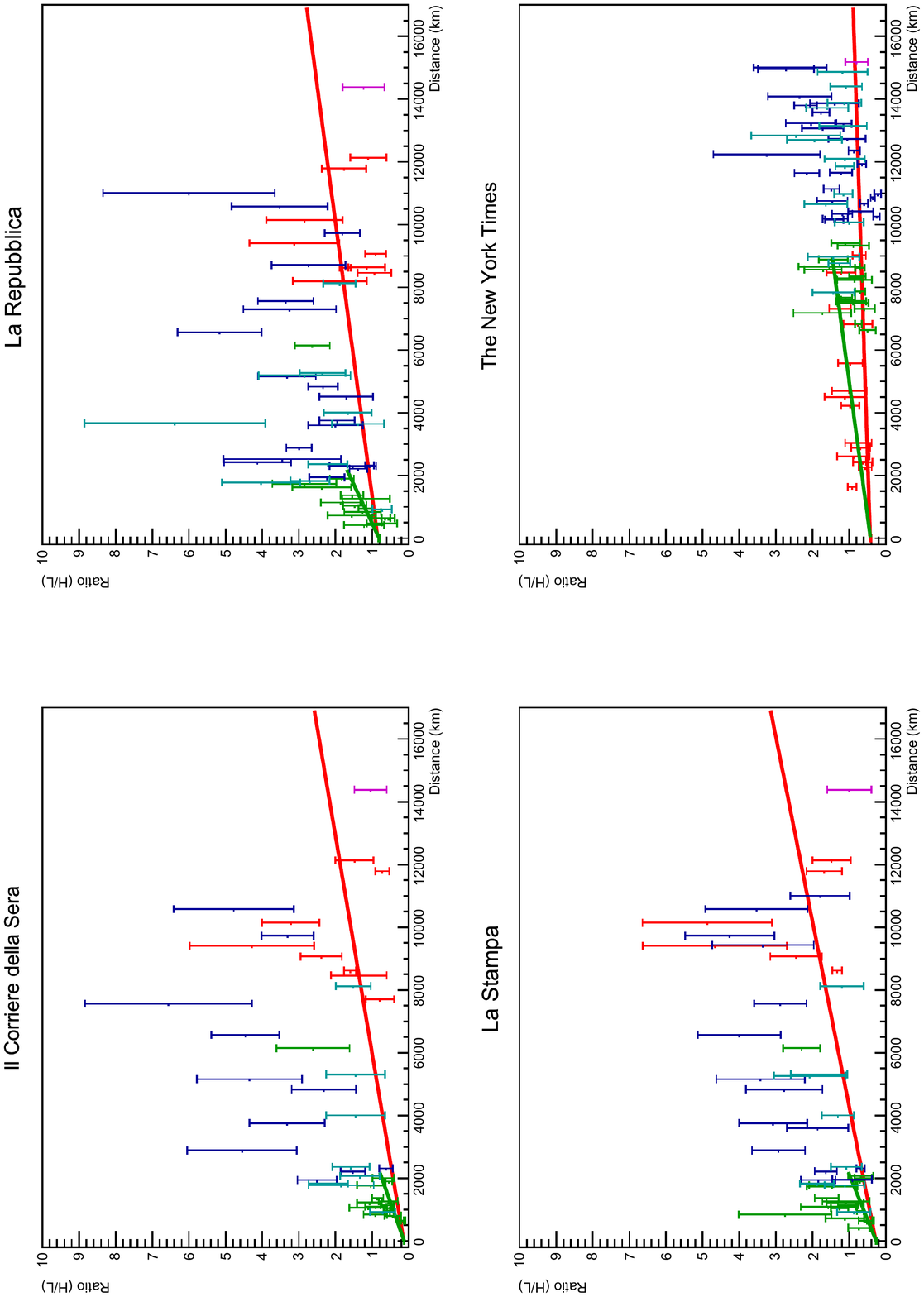}
 \caption{Distribution of the ratio $W_i=N_H/N_L$ vs distance between Rome (for CDS, REP, STA) and Washington (for NYT) and the capital of world nation $i$ with more than 30 events in each newspaper  dataset. Color denotes the continents: Green - Europe, Bue: Africa, Cyan: Asia, Red: America,  Purple: Oceania. Red lines indicate linear fit on the whole world, Green lines indicate linear fit only on Europe/USA.   }
\label{distratio}
\end{center}
\end{figure}

In Figure  \ref{distratio},  $W_i$ is plotted as a function of the distance $D_i$ between the capital of the country $i$       and Rome/Washington, according to the newspaper. It is possible to see how the value of $W$ tends to increase with the distance (fewer events with low $k$ and more with high $k$). The geographical bias appears to be stronger in Italian newspapers: a linear fit (Table \ref{tablinfit}) shows that the slopes  are similar in Italian newspapers and   higher by a factor $\simeq 5$ compared to NYT, a sign of an higher internationalization of the US paper: $dW/dx_{CDS}=(14\pm 1) \% / 1000\: km$; $dW/dx_{REP}=(11\pm 1) \% / 1000\: km$ and $dW/dx_{STA}=(17 \pm 1) \% / 1000\: km$, $dW/dx_{NYT}=(2.7\pm 0.3)\% / 1000\: km$. 

Social proximity effects also play a role in defining the values of the various countries: 
as also visible in Figure \ref{distratio}, American  countries have lower $W$ than equally distant Asian and   African ones.
    
If we limit the fit to countries in Europe (for Italian newspapers) and in America (for NYT) we find that the slopes are higher by a factor 2 to 4 than for those considering all  world nations (Table \ref{tablinfit}) a sign that geographical distance plays a stronger role for countries that are socially closed to either Italy or US (although the different geography of the American continent plays a role in the different behaviour).

\begin{table}[t]
\begin{center}
\begin{tabular}{cccc}
\hline
Newspaper 	&  All World &  Europe  / America  \\
 & $\delta W\: (\% / 1000 km)$ & $\delta W\: (\% / 1000 km)$ \\
\hline
NYT &2.70$\pm $0.3&  12$\pm $2  \\
CDS &14.4$\pm $0.9&  30$\pm $3  \\
REP &11.5$\pm $1&  40$\pm $9  \\
STA  &17$\pm $1&  35$\pm $4  \\
\hline
\end{tabular}\caption{Slope of the linear fit of the ratio $N_{2\leq k\leq 10}/N_{k>10} $ as a function of the distance between Rome/Washington and the various world countries. }
\label{tablinfit}
\end{center}
\end{table}

\subsection{Editorial rounding by excess of casualties as an additional  tool to detect censorship}

\begin{figure}[!ht]
\begin{center}
\includegraphics[width=.8\textwidth]{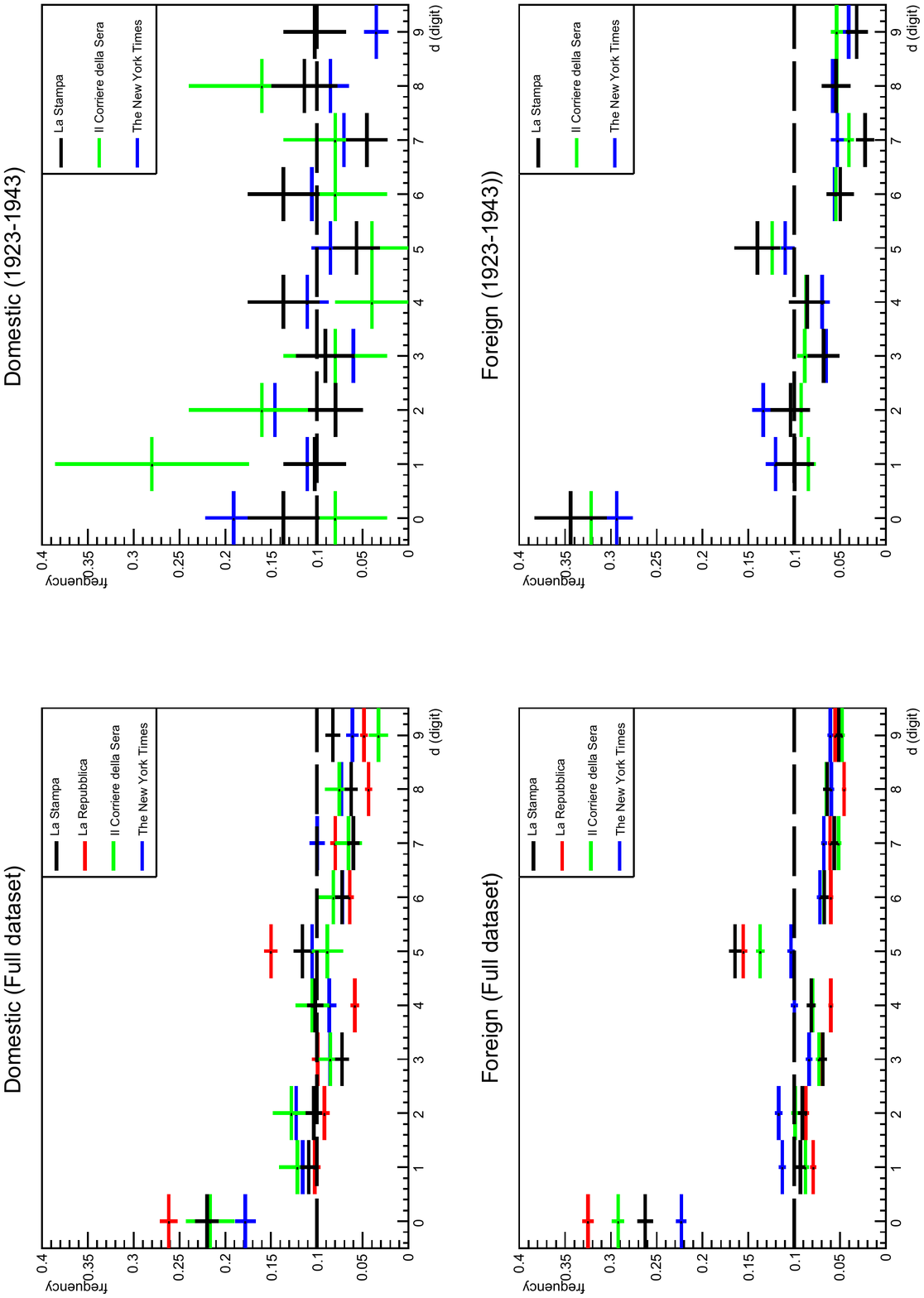}
  \caption{Relative occurrence of least significant digit for $10<k\leq 100$. Top: domestic, Bottom: foreign events. Left: full dataset, Right:  period of Fascist government in Italy. The dashed line shows the expected value from a random distribution (0.1). The values of `1' and `2' are closer to this value, with some defect in Italian newspapers and excess in NYT.  `3' and `4' are below expectations and their defect  is close to the excess of `5' events. The least probable digits are `6' to `9', usually rounded  in excess to `0'.  Only domestic events during Fascist government in Italy are not incompatible with equiprobable distribution.
 }
\label{digitsx}
\end{center}
\end{figure}

Newspapers often round up the number of casualties reported: this can be due to lack of knowledge,  to simplify  the headline, or to purposely increase emphasis  to attract the attention of the reader.   In the absence of tampering, we would expect the  least significant digit of $k$, $l_k$ to follow a uniform  distribution ($P(l_k)=1/10$). However, in Figure \ref{digitsx}, which shows the distribution $l_k$ for $10<k\leq 100$, it is possible to see how the values `0' and `5'   are overabundant (`5' only in the Italian papers) and the others under-abundant in respect to the flat distribution expectation. The number of defects in the digits `6' to `9' are close to the excess of  `0' (and similarly for `1' to `4' with `5'), confirming  the artificial nature of the reported number of casualties.
 Overall, Italian newspapers have a value of $l_0+l_5= 40-46\%$  and NYT has a  value of 32$\%$ (with respect to the 20$\%$ expected for the sum of the $k_0+k_5$  bins).
All distributions (see Table T1 in the suppl. mat.) are incompatible with the null hypothesis of the flat distribution with $p>0.01$. 

This distribution   is found in all newspapers and historical periods with one notable exception: during the Fascist government, the domestic distributions of STA and CDS   do not exhibit the peaks for $k_0$ and $k_5$, still present in the corresponding foreign distributions of the same period. A $\chi^2$ test allows us to reject the hypothesis that foreign and domestic histograms of Italian newspapers  follow the same distributions with $p>0.01$. Furthermore, the domestic  distributions of STA and CDS are the only ones that are not incompatible with the equiprobable one  (Figure S3  and Table T1 in the Supplementary Information).
This implies that during the Fascist regime the editorial practice was to  report more faithfully the number of domestic casualties as an additional way of suppressing these events   and  at the same time exaggerating the number of foreign casualties.   

This phenomenon is similar but specular to Benford's law\cite{Newcomb, benford193810.2307/984802}, which describes the  statistical distribution  of the most significant digit in several natural and man-made datasets\cite{Morzy2016}. Benford's law has been used to determine the presence of accounting\cite{benfordaccounting} or election fraud\cite{Jimeneze1602363}, since values that are artificially altered do not follow it.

\section{Conclusions}

 In this work we have developed a series of techniques to  automatically scan the complete historical  databases of daily newspapers for the occurrence of specific keywords related to accidents and death. 
We also devised various    algorithms to analyze the dependence of these keywords from the geographical location and historical period.
Over traditional analysis, consisting of manual scanning of the newspaper articles, these tools offer the advantage of being  automatic and thus being applicable on larger datasets spanning longer time periods.  
Indeed, these tools are suited for historical analysis to evaluate quantitatively the presence of bias or censorship in a given publication and its variation over time and political environment. This paper  considered printed daily newspapers, but these techniques can  be used also on online publications, news outlets, etc... 
Although the usage is limited only to articles with quantitative keywords, where to the  event (accident, casualty, death...) is associated the number of persons involved (The simpler word occurrence methods  can be used on a wider word set), this type of analysis is complementary to the assessment of  'fake news' since has the advantage of being automatic and operating on large data sets.  These tools can also contribute to the assessment of the  freedom of the press in a  given country.

Future work will extend the application of these tools to other lemmas such as {\it casualties, wounded, victims}. The analysis will be applied to differences in  reporting between the type of accidents, both of man-made origin (e.g. train/airplane/ship) and natural calamities.  Also the structure of reporting as a function of the day of the week and the page location will be considered. On a larger scope. also other  newspapers, magazines,   online publications  will be considered, extending the analysis to look  for the presence and evolution of   ethnic or national bias. A more long-term goal can also be the use of speech recognition methods to study the occurrence of these lemmas on radio and TV.

\section{Abbreviations}
\begin{itemize}
\item CDS - Il Corriere della Sera
\item NYT - The New York Times
\item OCR - Optical Character Recognition
\item REP - La Repubblica
\item STA - La Stampa
\item WW1 - World War 1
\item WW2 - World War 2
\end{itemize}

\section{Declarations}
\subsection{ Availability of data and material}
The newspaper archives are accessible at these locations:

{\it New York Times}: https://www.nytimes.com/search/

{\it Il Corriere della Sera} http://archivio.corriere.it/ 

{\it La Repubblica}: https://ricerca.repubblica.it/ 

{\it La Stampa}: http://www.archiviolastampa.it/

\subsection{Acknowledgments}
 The author is grateful to  Dr. A. Truzzi   for the useful and stimulating discussions during the preparation of this work and  Drs U. T. Casolino, W. Husein, O. Larsson, L. Marcelli, L. Sorge and M. Piersanti for reviewing the manuscript.
  
The author also wishes to thank the four newspapers considered ({\it The New York Times, Il Corriere della Sera, La Repubblica, La Stampa})  for having their historical archives freely available for consultation: without these resources the paper could not have been written. 

\section{Appendix. Estimation of the missing events}
\label{mathappendix}

For a power-law distribution of the number of articles $N(k)$ reporting that $k$ people have been killed:
\begin{equation}
N(k)=\alpha k^{-\gamma} 
\end{equation} 

the total number of articles with two or more people killed is:
\begin{equation}
N_{tot}=\frac{\alpha}{\gamma-1} 2^{1-\gamma}  \:\:\:\:\: \gamma>1
\end{equation} 
 
for a given $N_{tot}$ we have therefore that:

\begin{equation}
N(k)=\frac{N_{tot}(\gamma-1)}{2^{1-\gamma}} k^{-\gamma} 
\end{equation}

\begin{figure}[!ht]
\begin{center}
\includegraphics[width=\textwidth]{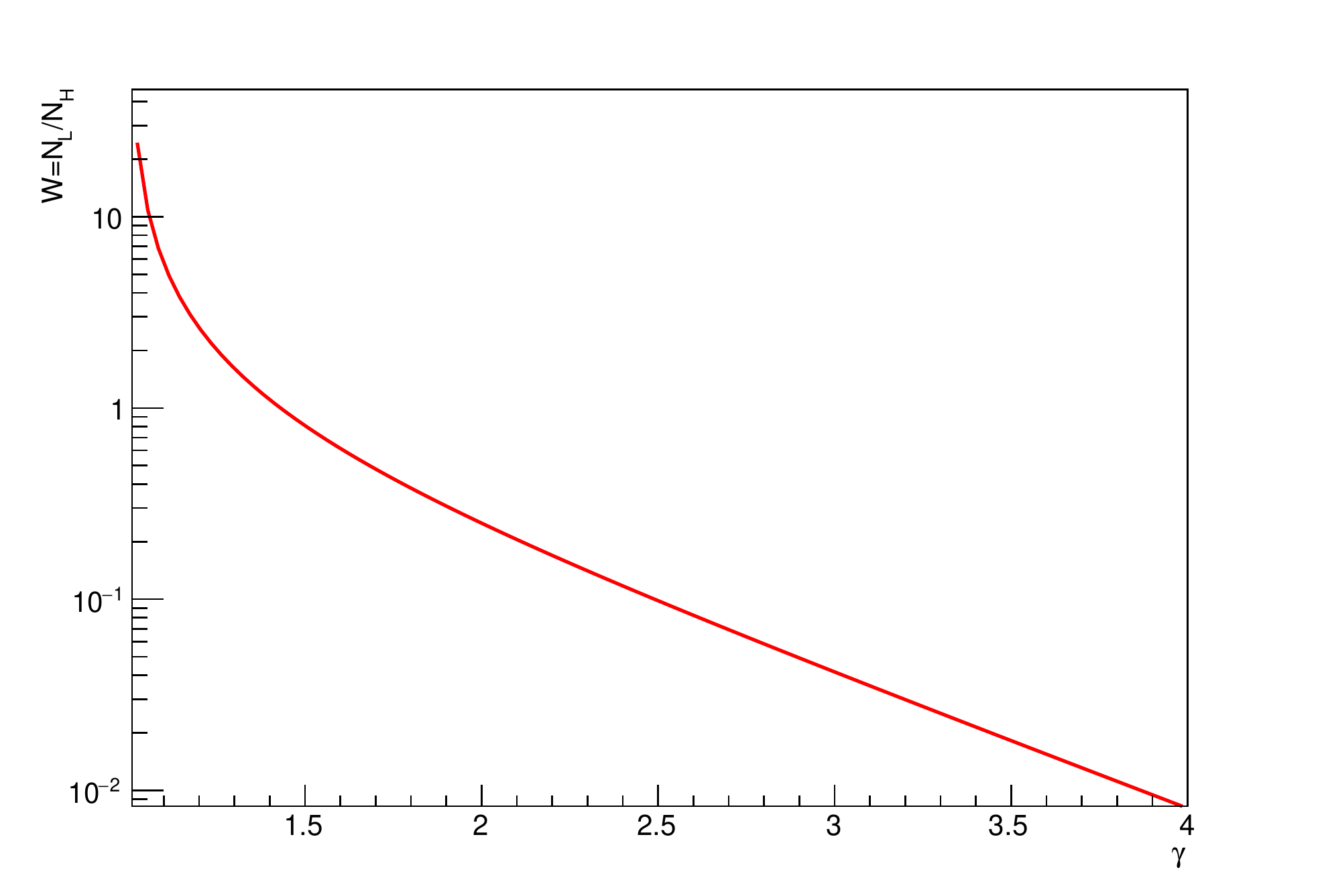}
\caption{Ratio $W=N_H/N_L$ as a function of $\gamma $ for a single power-law distribution of $N_k$. A value of $\gamma=1.43$
results in $W=1$, an equal number of articles with $2\leq k\leq 10$ and $k>10$.   }
\label{theor_ratio}
\end{center}
\end{figure}

In a single power law, the ratio $W$ of articles with less ($N_L$) and more ($N_H$) than 10 people dead, is:
  \begin{equation}
 W = \frac{10^{1-\gamma}}{2^{1-\gamma}-10^{1-\gamma}}=\frac{1}{5^{\gamma-1}-1}
\end{equation}
A high value of $\gamma$ implies an emphasis on articles with low $k$ and vice-versa. For $\gamma=1.43$
we have  an equal number of articles with $2\leq k\leq 10$ and $k>10$ (Figure \ref{theor_ratio}).

For a broken power law distribution:
\begin{equation}
 N(k) = \left\{
\begin{array}{l l}
 \alpha k^{-\gamma_L} & 2\leq k \leq 10 \\
 \beta k^{-\gamma_H} & k>10  \\
\end{array} 
\right.
\end{equation}
 
so

\begin{equation}
 N_{L} = \frac{\alpha}{1-\gamma_L}\left( 10^{1-\gamma_L}-2^{1-\gamma_L}\right) \:\:\: 2\leq k \leq 10
\end{equation}
\begin{equation}
 N_{H} = -\frac{-\beta}{1-\gamma_H}\left. 10^{1-\gamma_H}\right. \:\:\:  k > 10
\end{equation}

\begin{equation}
M=1-\frac{  \int_{2}^{10}\beta k^{-\gamma_{H}} dk  }{\int_{2}^{10}\alpha  k^{-\gamma_{L}} dk }= 1- \frac{1-\gamma_{L}}{1-\gamma_{H}}10^{\gamma_{H}-\gamma_{L}}\frac{10^{1-\gamma_{H}}-2^{1-\gamma_{H}}}{10^{1-\gamma_{L}}-2^{1-\gamma_{L}}}
\end{equation}

\begin{figure}[!ht]
\begin{center}
\includegraphics[width=\textwidth]{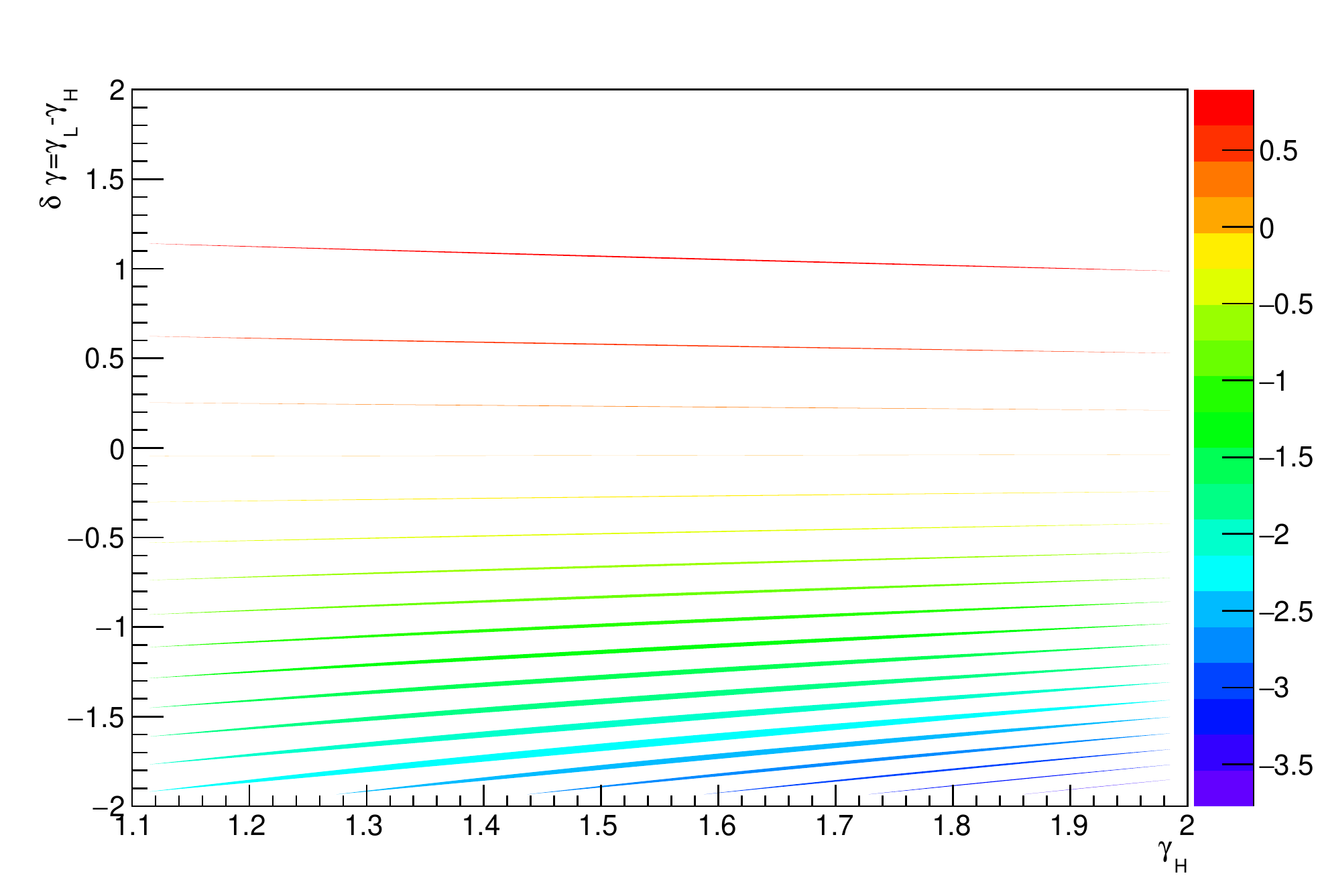}
\caption{$M$, amount of lack ($M<0$) or excess ($M>0$) articles in respect to a pure power law distribution ($\delta_\gamma=0$).  }
\label{missing}
\end{center}
\end{figure}

The excess ($M>0$)  or defect ($M>0$) in respect to a pure power law ($M=0$) thus depends on the two values $\gamma_H$ and $\gamma_L$. They are shown in Figure \ref{missing}.

\begin{figure}[!htbp]
\begin{center}
 \includegraphics[width=0.42\textwidth,angle=-90]{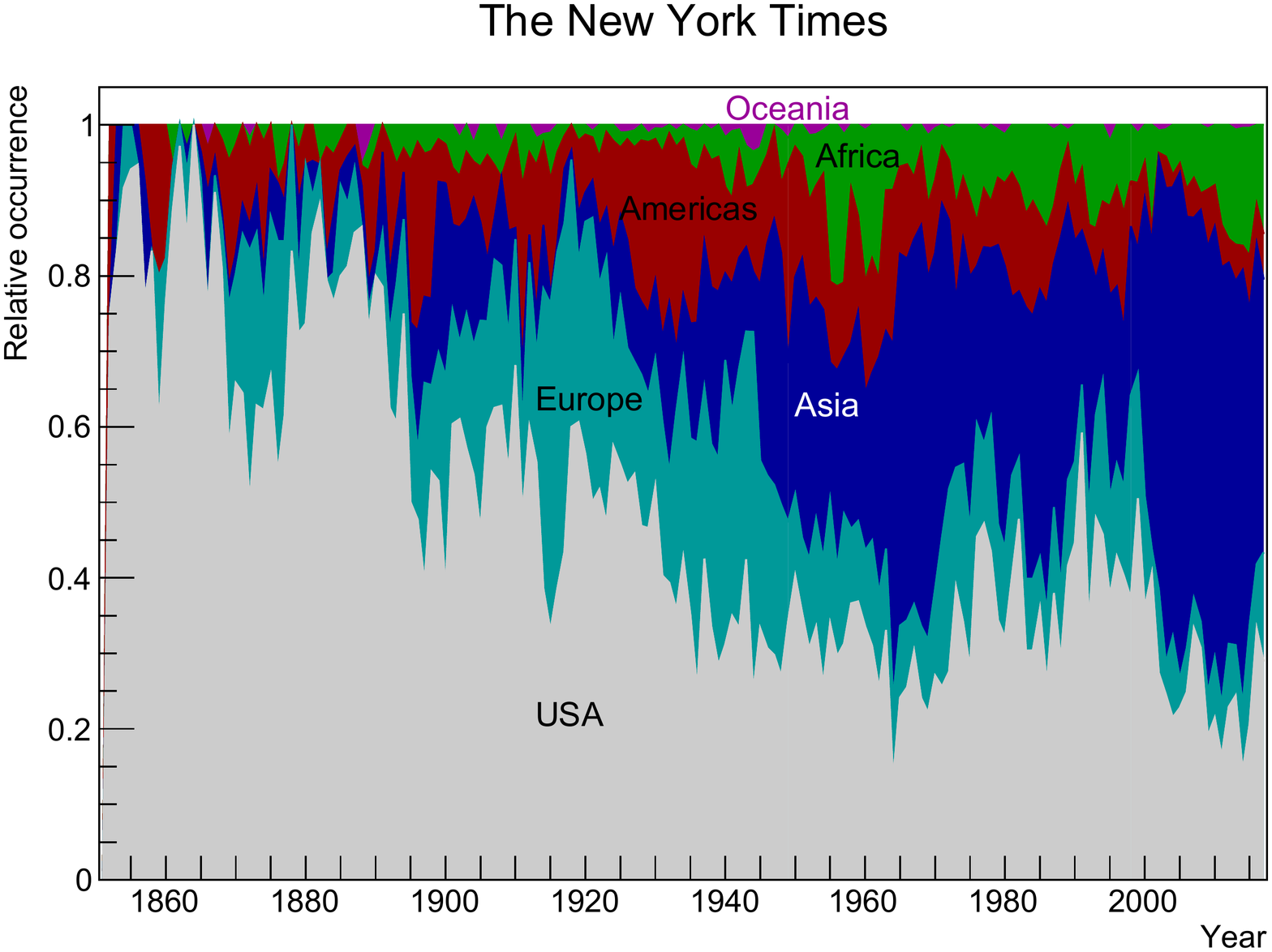}
 \includegraphics[width=0.42\textwidth,angle=-90]{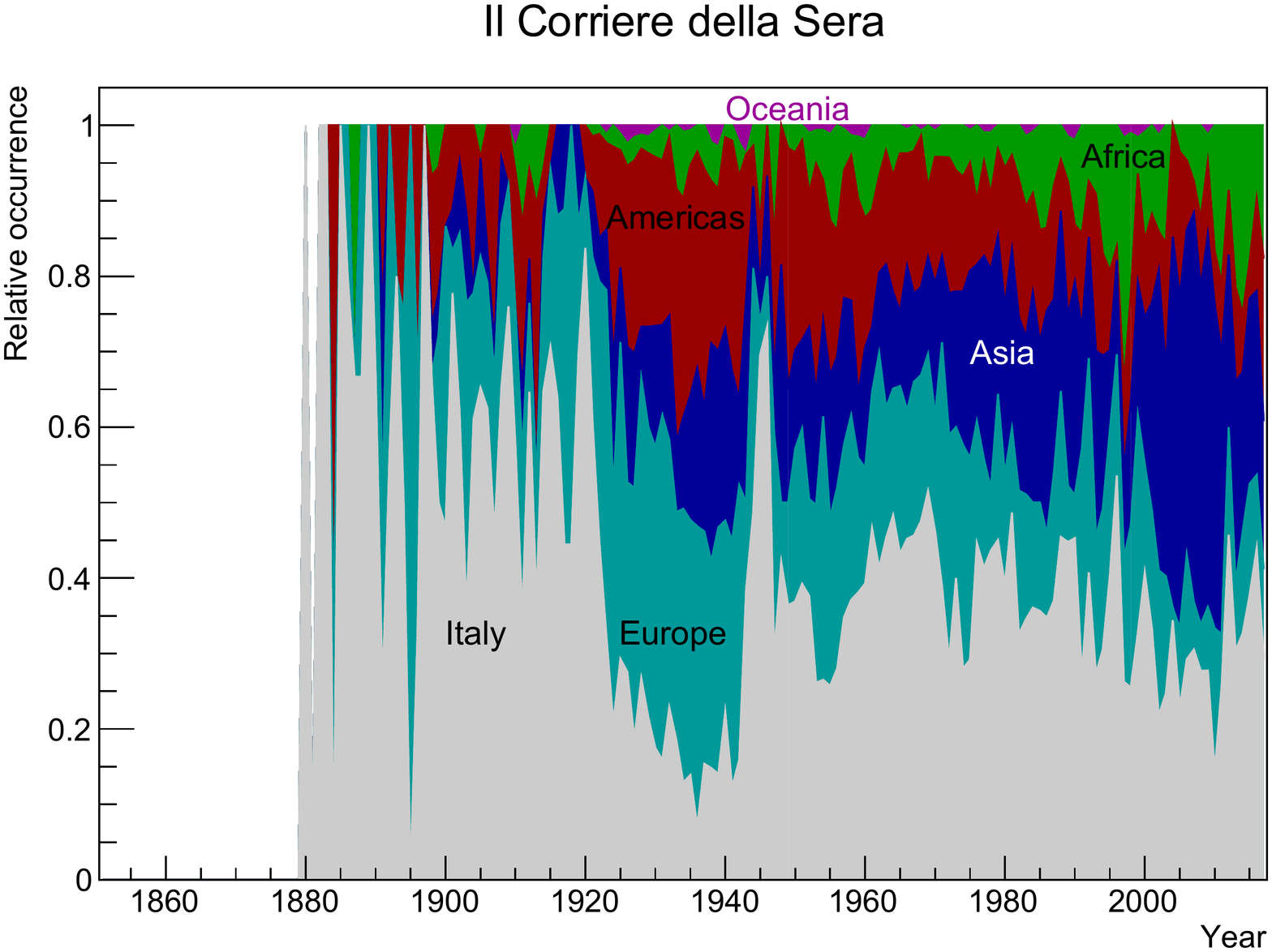}
 \\
 \includegraphics[width=0.42\textwidth,angle=-90]{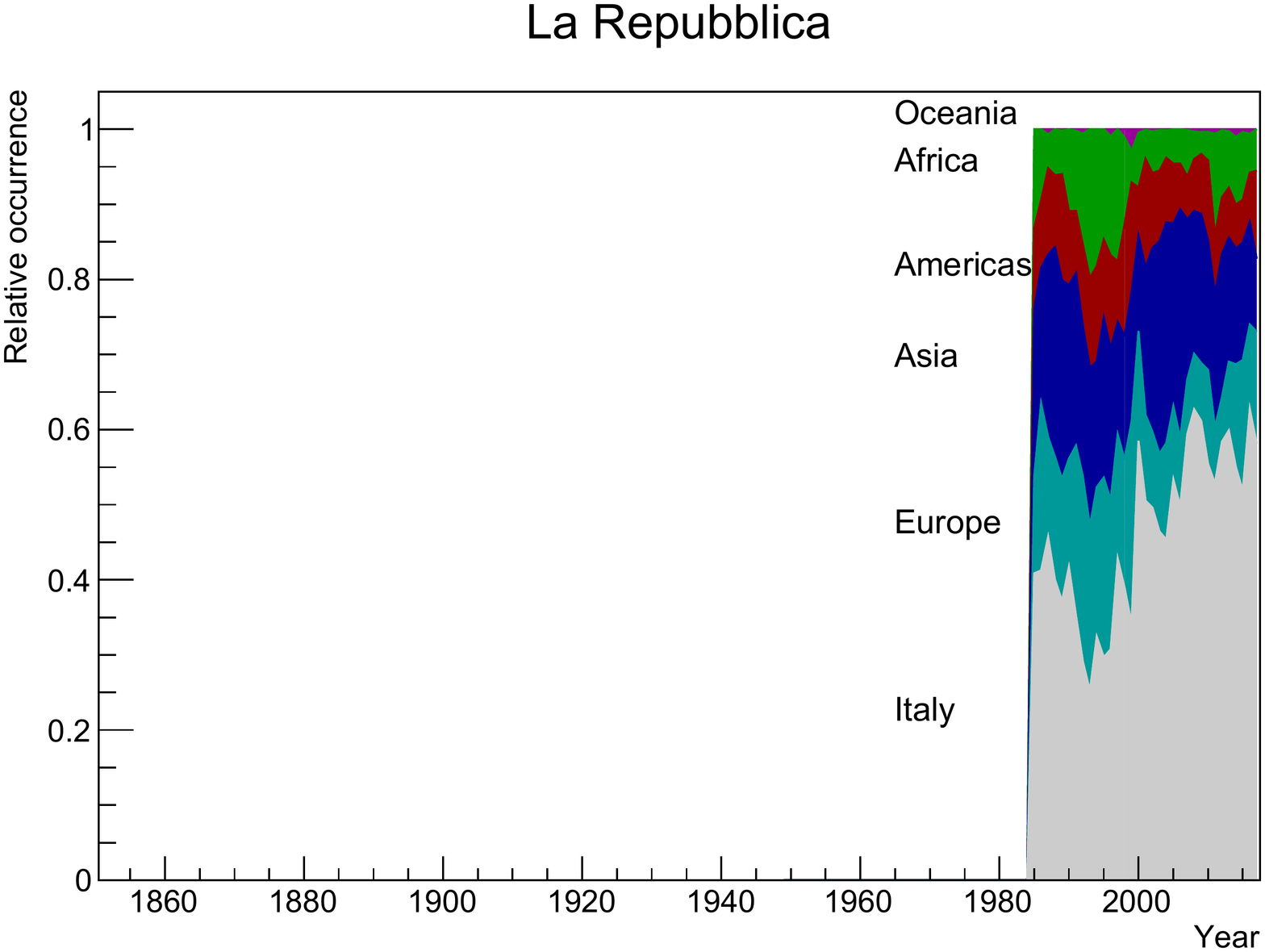}
 \includegraphics[width=0.42\textwidth,angle=-90]{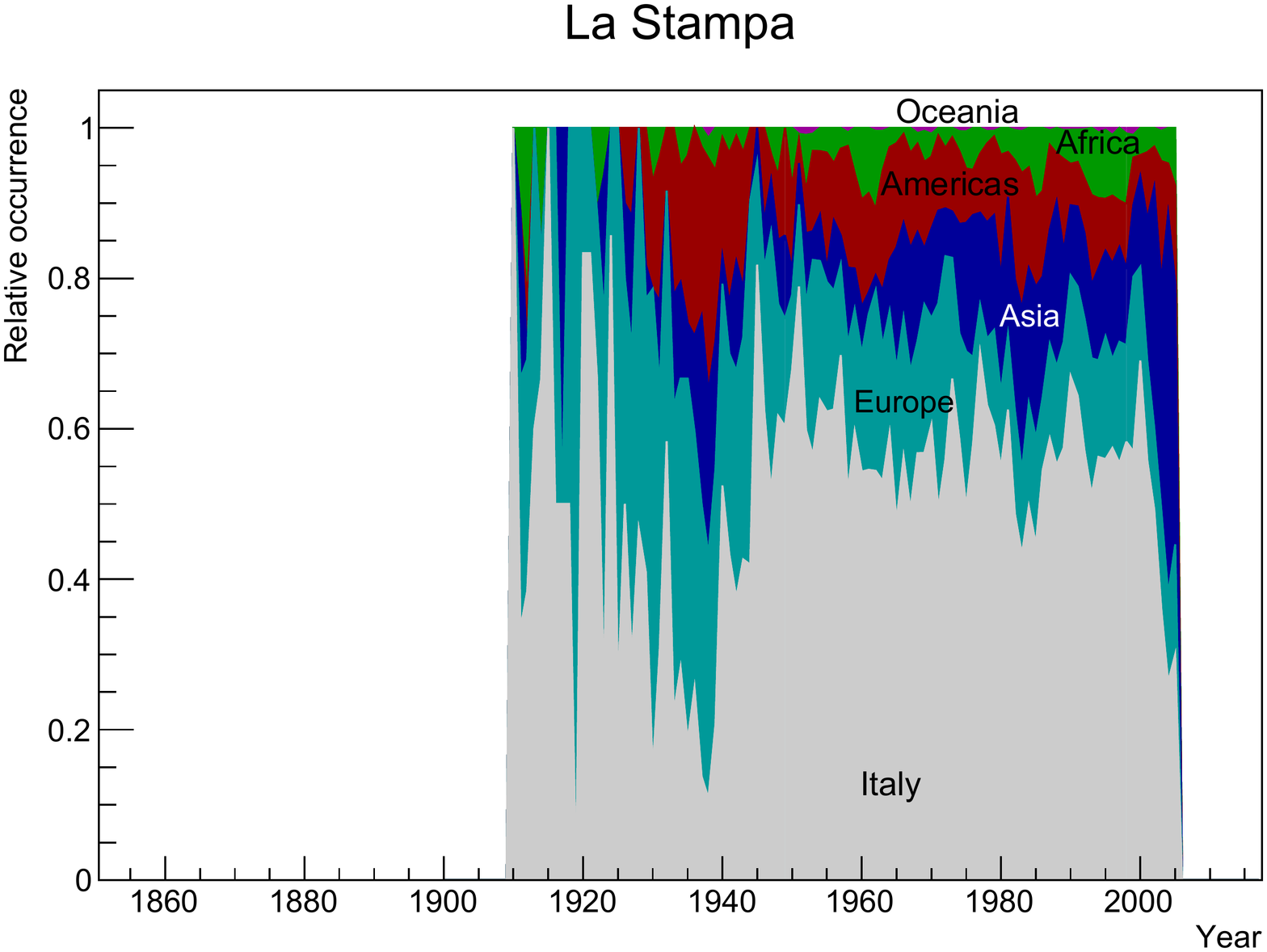}
 \caption{ Normalized ratio of articles as function of time for domestic and foreign deaths in the various continents. 
It is possible to see how in the NYT the percentage of domestic events decreases with time. After WW2, the percentage of Asian events increases rapidly. In CDS and STA it is possible to see  the decrease of domestic events during Fascism and the sharp peak after WW2.  }
\label{xxcontinentixx}
 \end{center}
\end{figure} 

\section{Supplementary online material}

\subsection{Systematic error associated to finite sample of newspapers}

To asses the systematic error arising from the finite sample of the events for a given selection we have performed the power law fitting on $N_{test}=1000$ different subsets, each obtained removing a  percentage $P_{cut}$ of random events from those passing the cut. 
Fitting has been performed on each of these distributions and the resulting values of $\gamma $ histogrammed.  
The sigma resulting from a  Gaussian fit of each histogram of values of $\gamma$  at a given $P_{cut}$ can be used as an estimation of the systematic error associated to the spectral index $\gamma$. 

\begin{figure}[!ht]
\begin{center}
\includegraphics[width=0.60\textwidth,page=21]{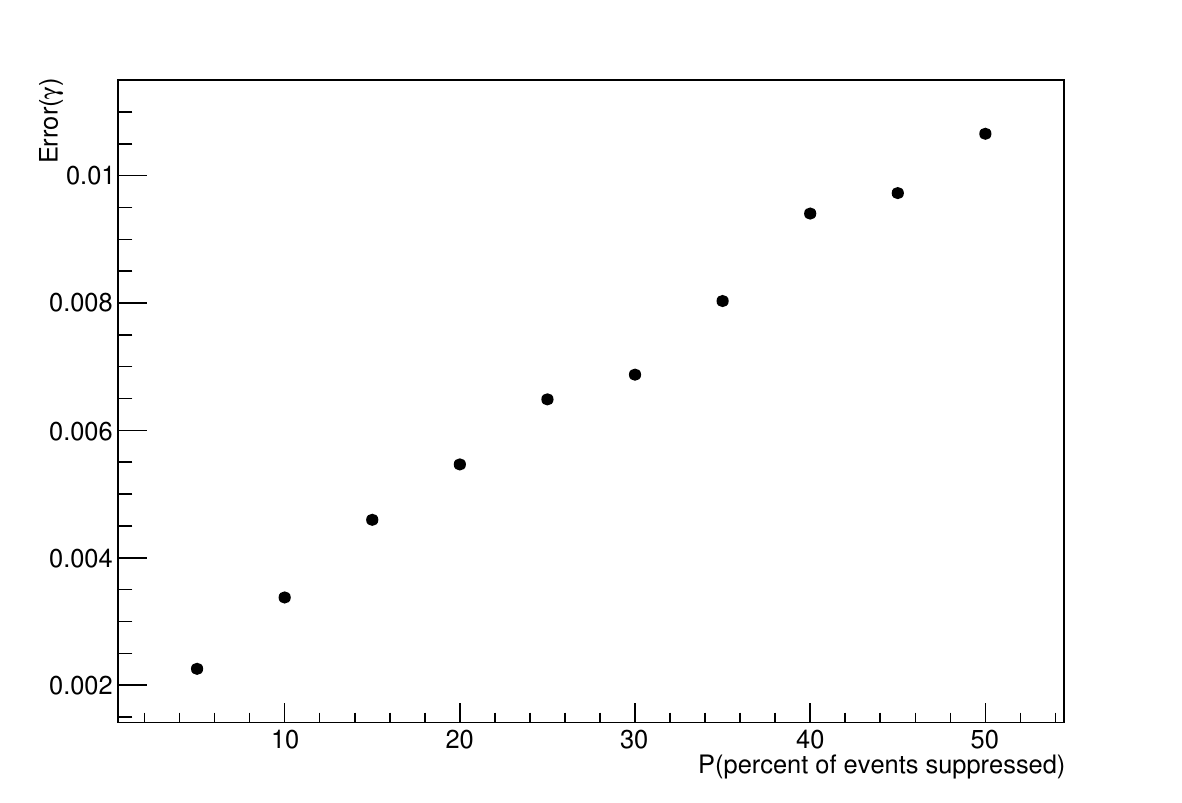}
  \caption{Sigma of gaussian fit of the histogram of the values of $\gamma $ as function of  $P_{cut}$. Full dataset of NYT. 
	}
\label{probfit3}
\end{center}
\end{figure}

Values of $5\leq P_{cut} \leq 50\% $  have been removed to test    the robustness of the algorithm to the finite data set. As expected, the  error grows more slowly  with the increase of $P_{cut}$ for larger samples. For instance, for the overall spectral index of the newspapers (see Figure \ref{probfit3}) this goes from 0.002 for $P_{cut}=5\%$  to  0.02 for $P_{cut}=50\%$. 
If we assume a very conservative value of $P_{cut}=30\%$ as incompleteness of the data set,  we can estimate the systematic error due  fitting to be 0.07 for the full dataset. 
Unless otherwise noted, this value has been added to the statistical errors in the plots. 

\subsection{Distributions of least significant digit}
\label{somdigit}

\begin{figure}[!ht]
\begin{center}
\includegraphics[width=0.48\textwidth,page=1]{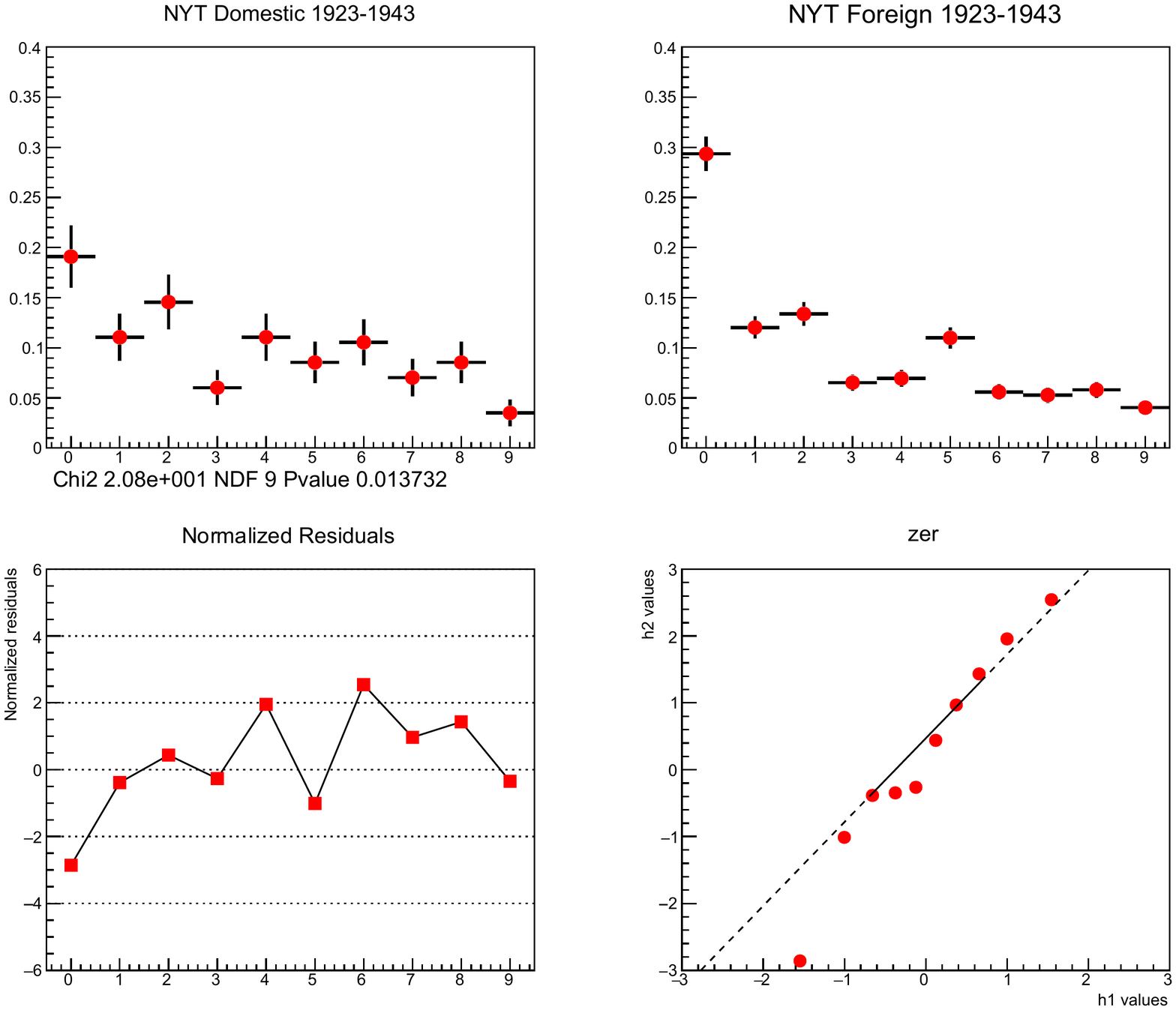}
\includegraphics[width=0.48\textwidth,page=2]{./spectraldigits_2.pdf}
\includegraphics[width=0.48\textwidth,page=3]{./spectraldigits_2.pdf}
\includegraphics[width=0.48\textwidth,page=4]{./spectraldigits_2.pdf}
\caption{$\chi^2$ test comparing various distributions of the least significant digit for $10<k\leq 100$. Top Left and right: the two histograms (or histo and flat distribution function) being compared. Bottom left: Normalized residuals, Bottom right: QQ plot. NYT.  
  } 
\label{digitsN}
\end{center}
\end{figure}

For NYT (Figure \ref{digitsN} shows the $\chi^2$ test comparing various  distributions of the least significant digit for NYT: domestic - foreign, domestic - flat, foreign - flat, all - flat (full dataset). From the resulting Chi2 and p value it is possible to see how none of the distributions are compatible with a flat hypothesis. The analysis of the residuals and the QQ plot show how the highest deviations occur in the digits corresponding to `0' and `5').

\begin{figure}[!ht]
\begin{center}
\includegraphics[width=0.48\textwidth,page=5]{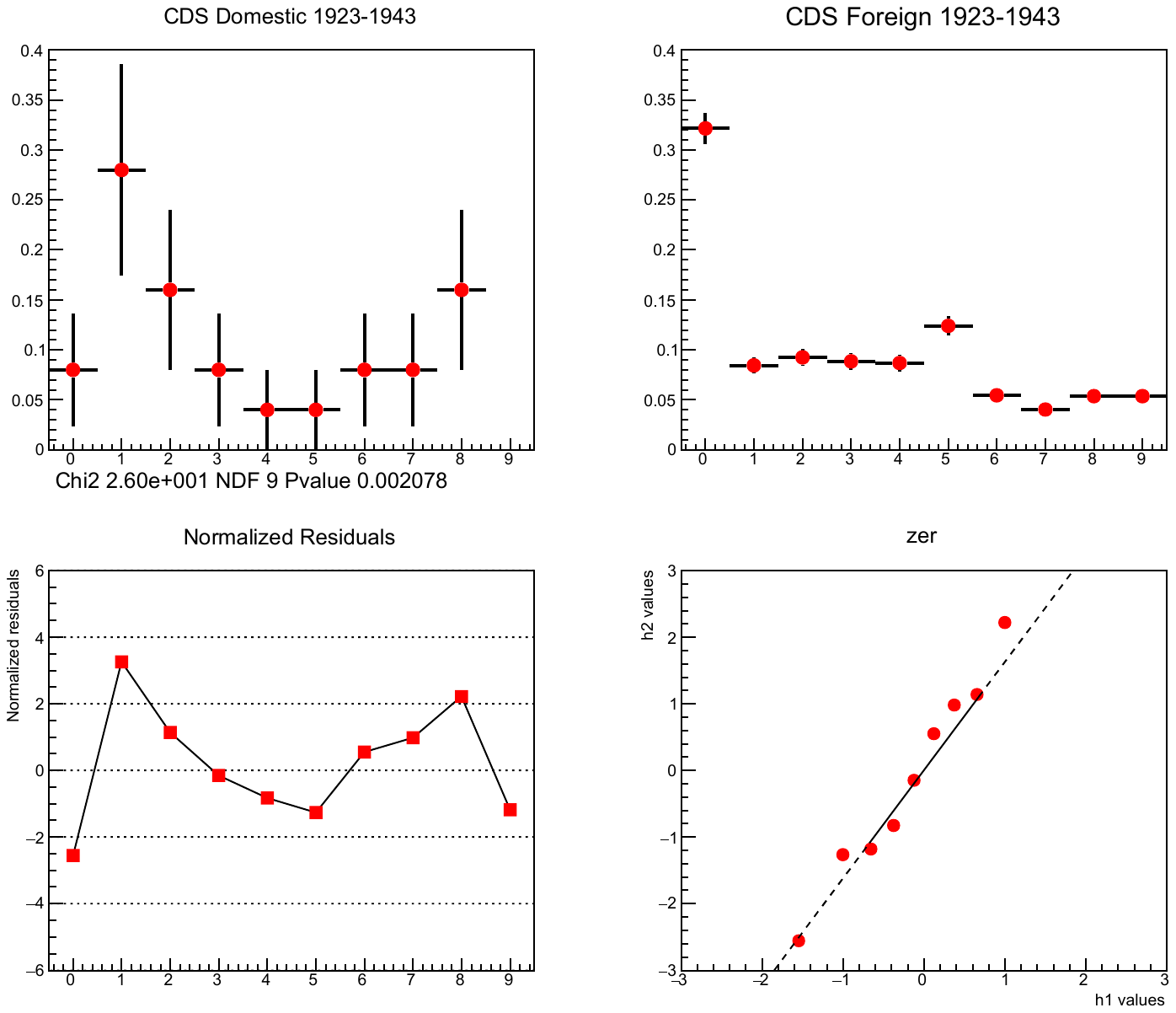}
\includegraphics[width=0.48\textwidth,page=6]{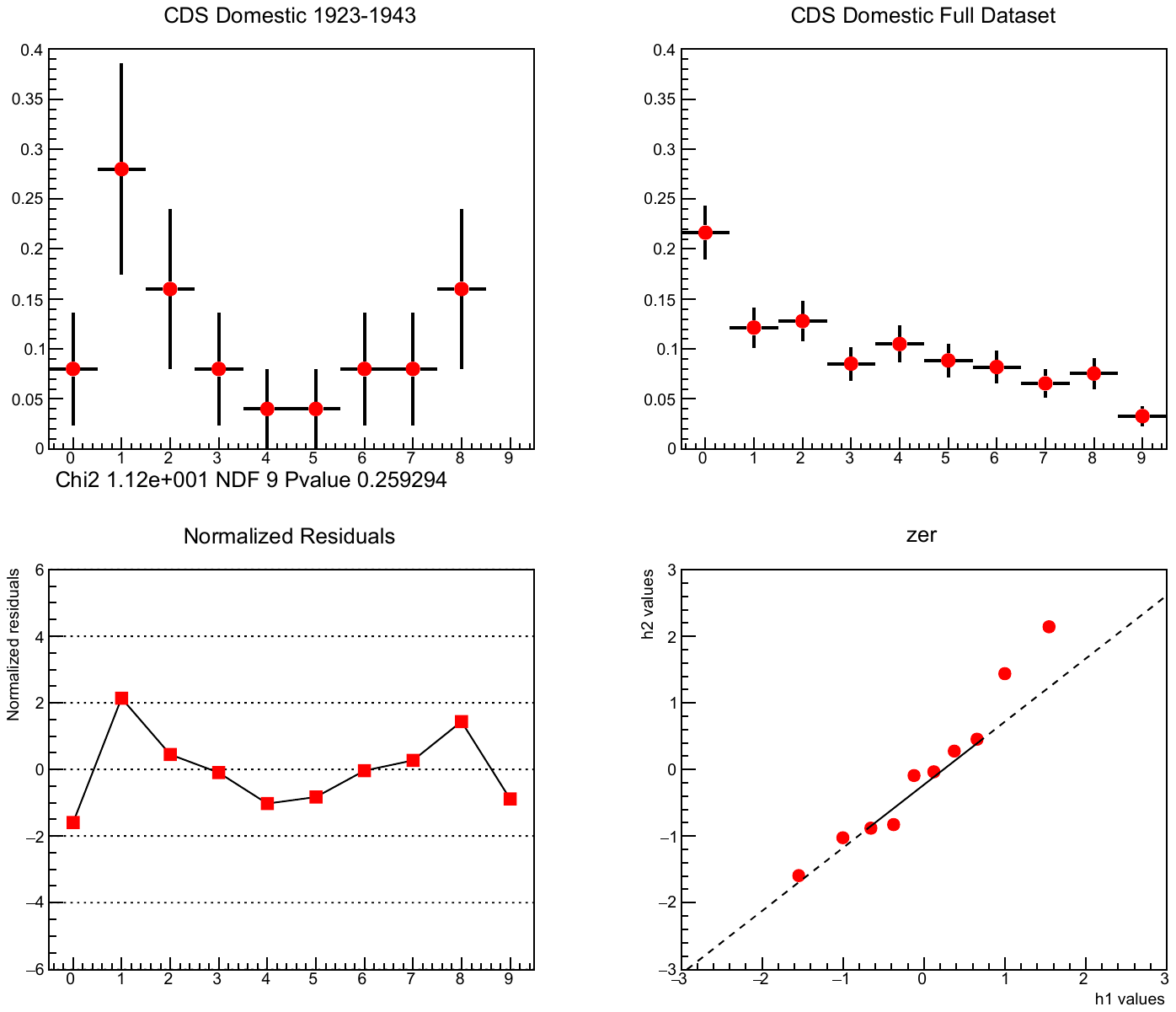}
\includegraphics[width=0.48\textwidth,page=7]{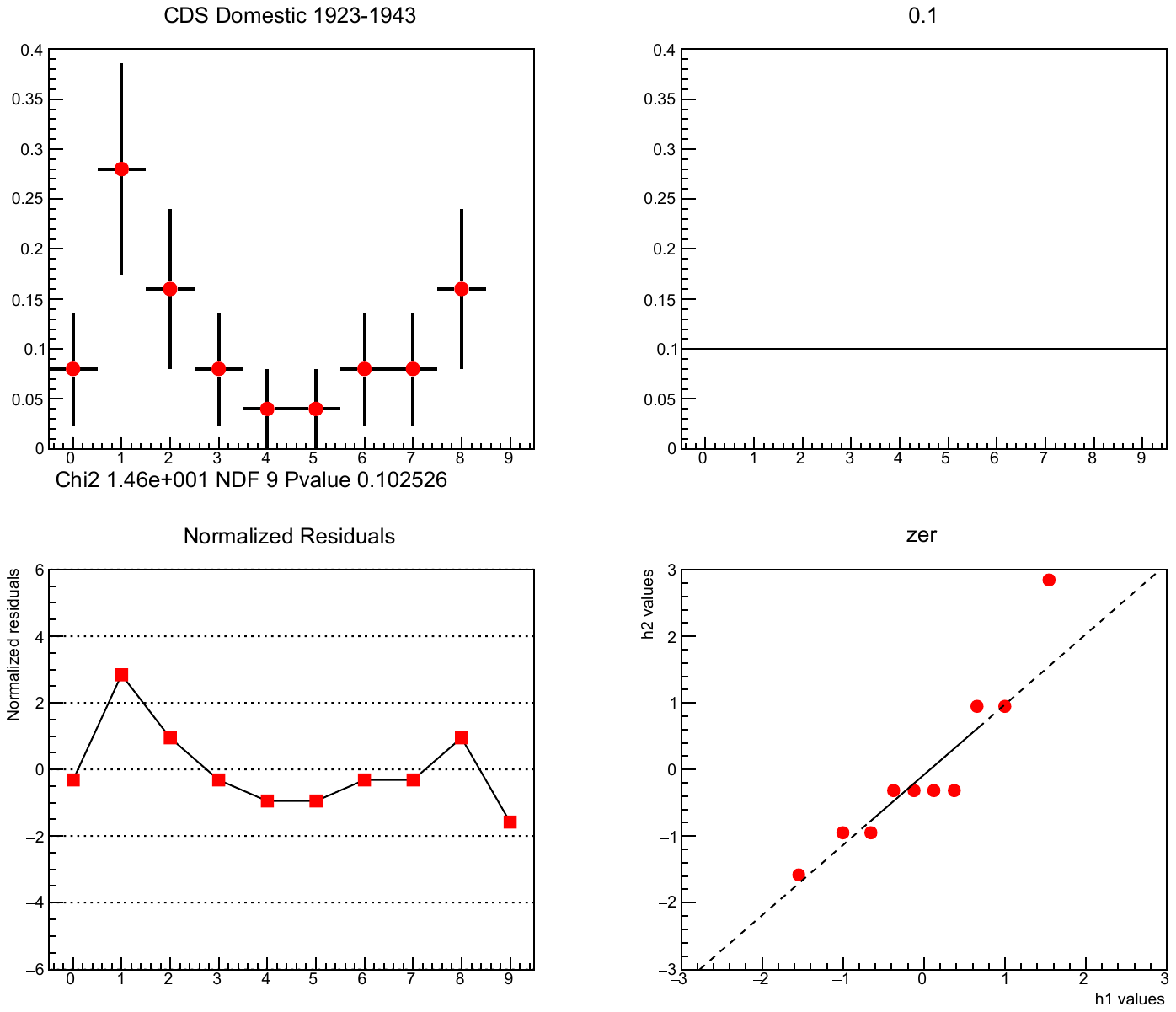}
\includegraphics[width=0.48\textwidth,page=8]{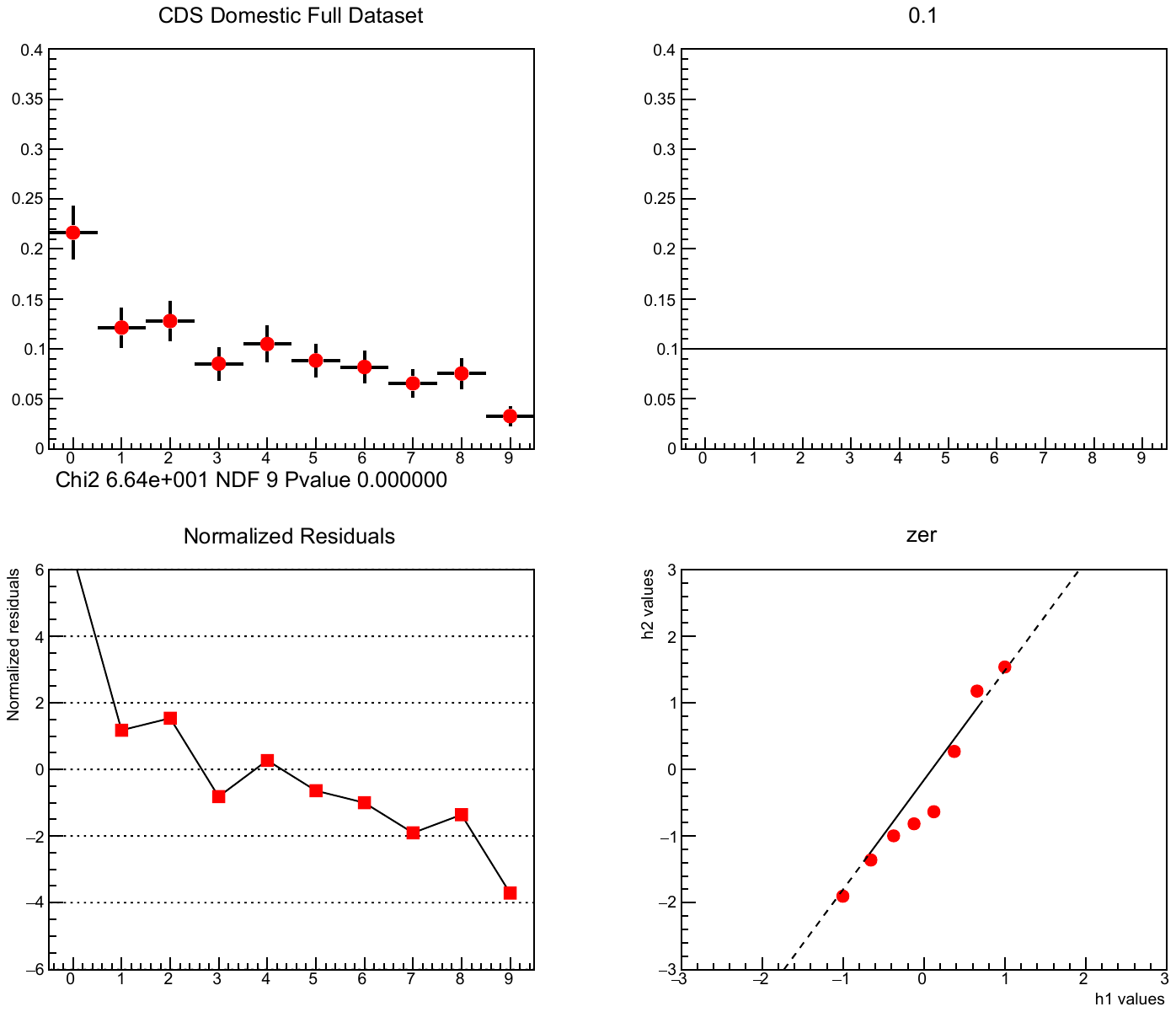}
\caption{$\chi^2$ test comparing various distributions of the least significant digit for $10<k\leq 100$. Top Left and right: the two histograms (or histo and flat distribution function) being compared. Bottom left: Normalized residuals, Bottom right: QQ plot. CDS.  
 }
\label{digitsC}
\end{center}
\end{figure}

\begin{figure}[!ht]
\begin{center}
\includegraphics[width=0.48\textwidth,page=9]{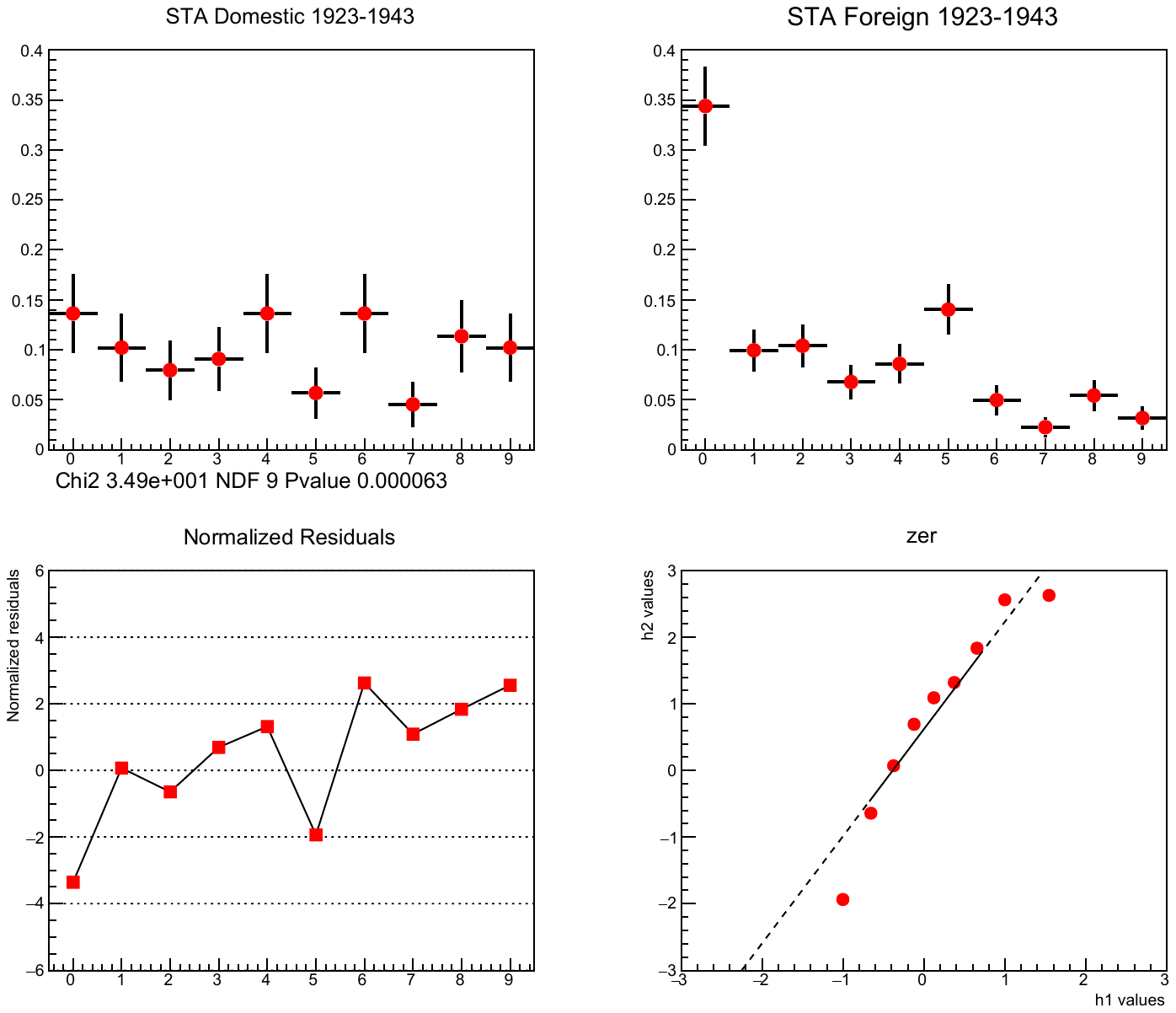}
\includegraphics[width=0.48\textwidth,page=10]{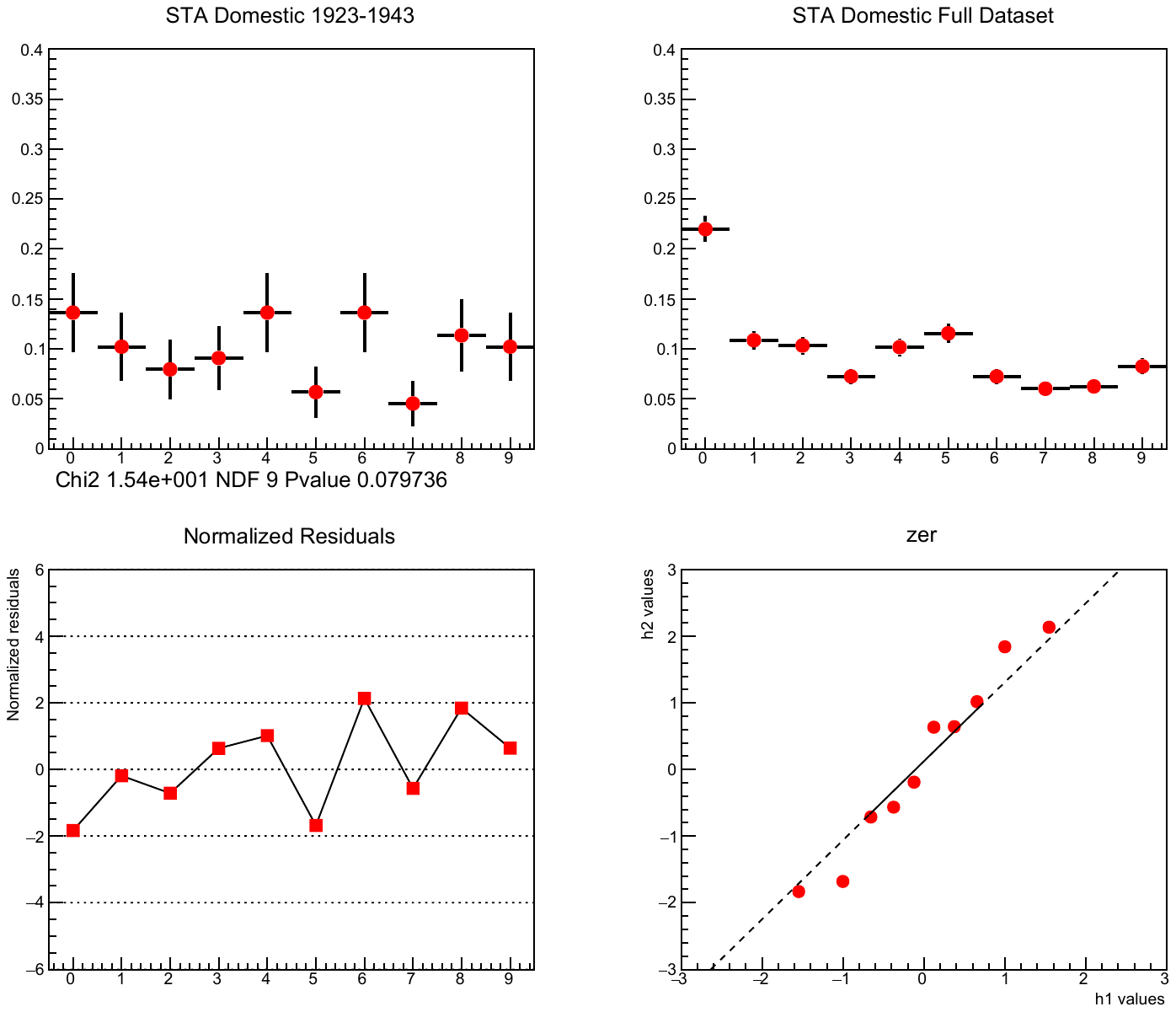}
\includegraphics[width=0.48\textwidth,page=11]{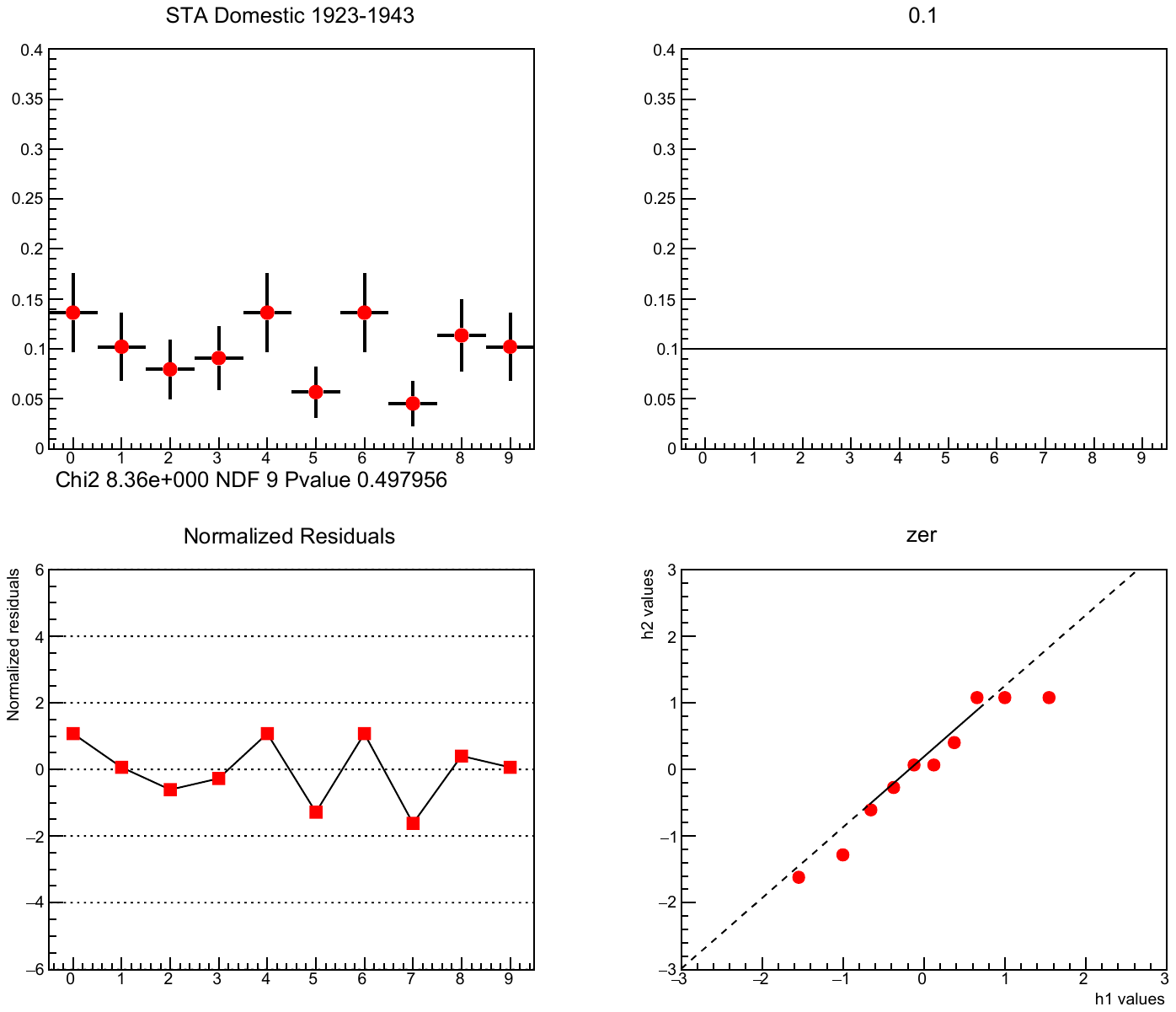}
\includegraphics[width=0.48\textwidth,page=12]{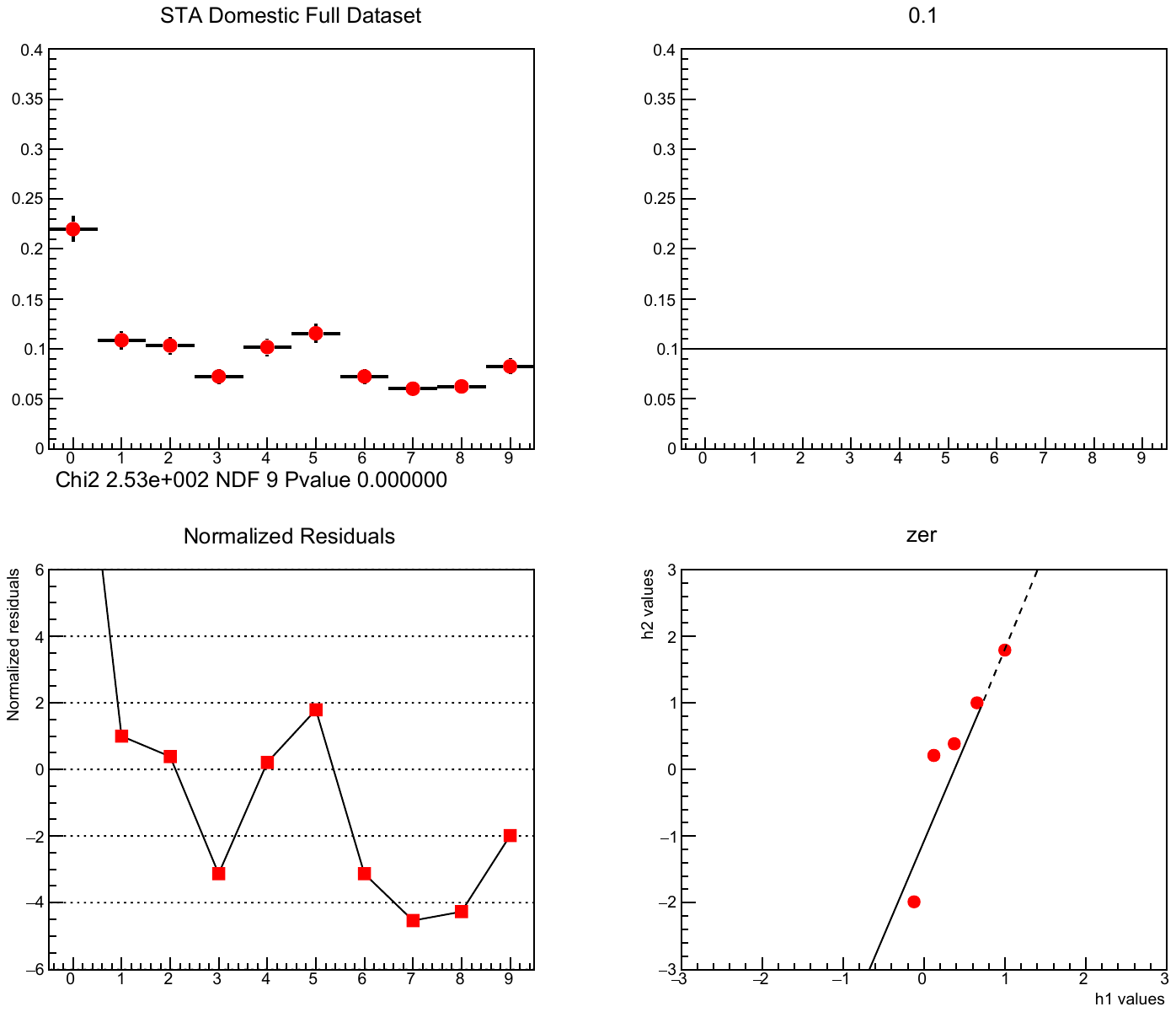}
\caption{$\chi^2$ test comparing various distributions of the least significant digit for $10<k\leq 100$. Top Left and right: the two histograms (or histo and flat distribution function) being compared. Bottom left: Normalized residuals, Bottom right: QQ plot. STA.  
}
\label{digitsS}
\end{center}
\end{figure}

\begin{figure}[!ht]
\begin{center}
\includegraphics[width=0.48\textwidth,page=13]{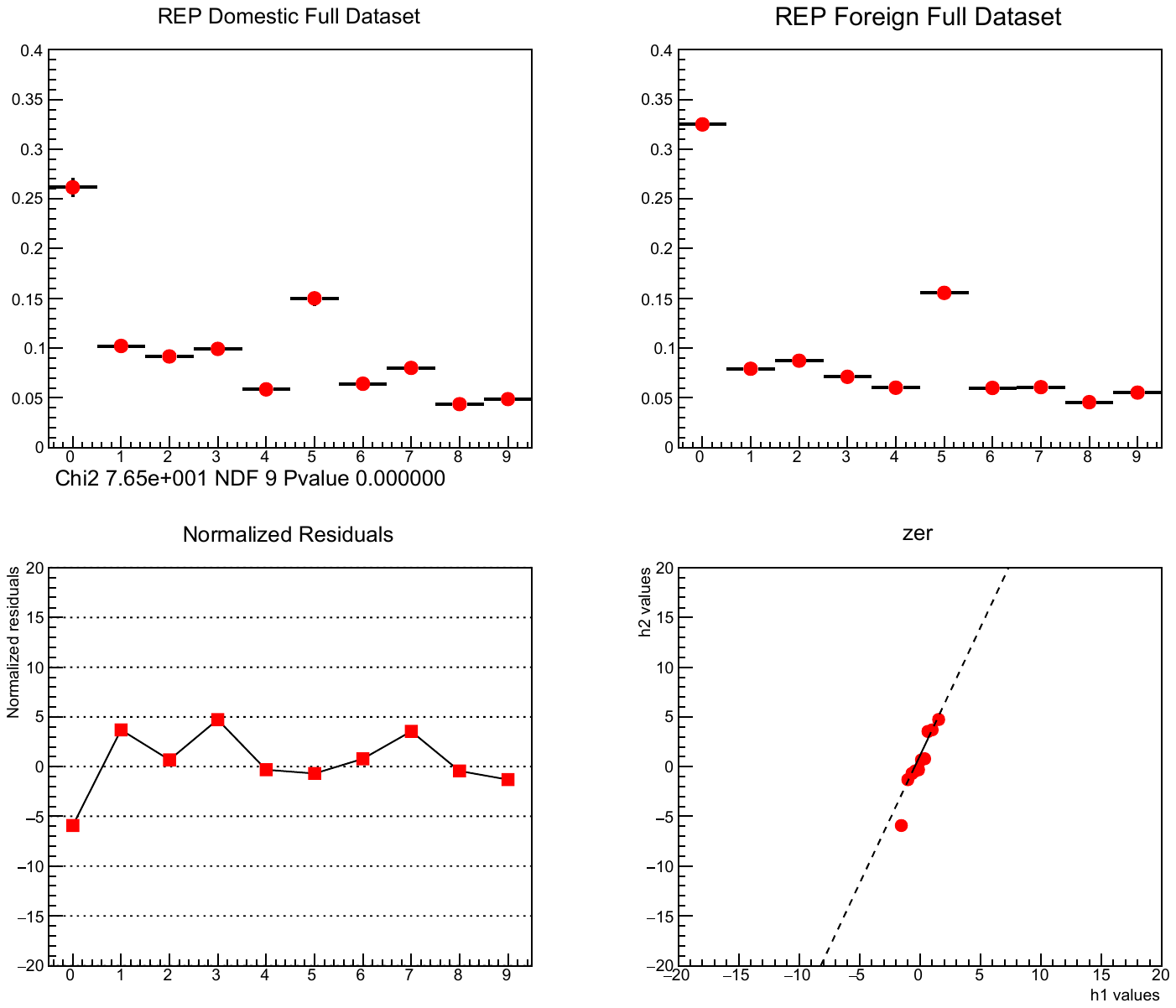}
\includegraphics[width=0.48\textwidth,page=14]{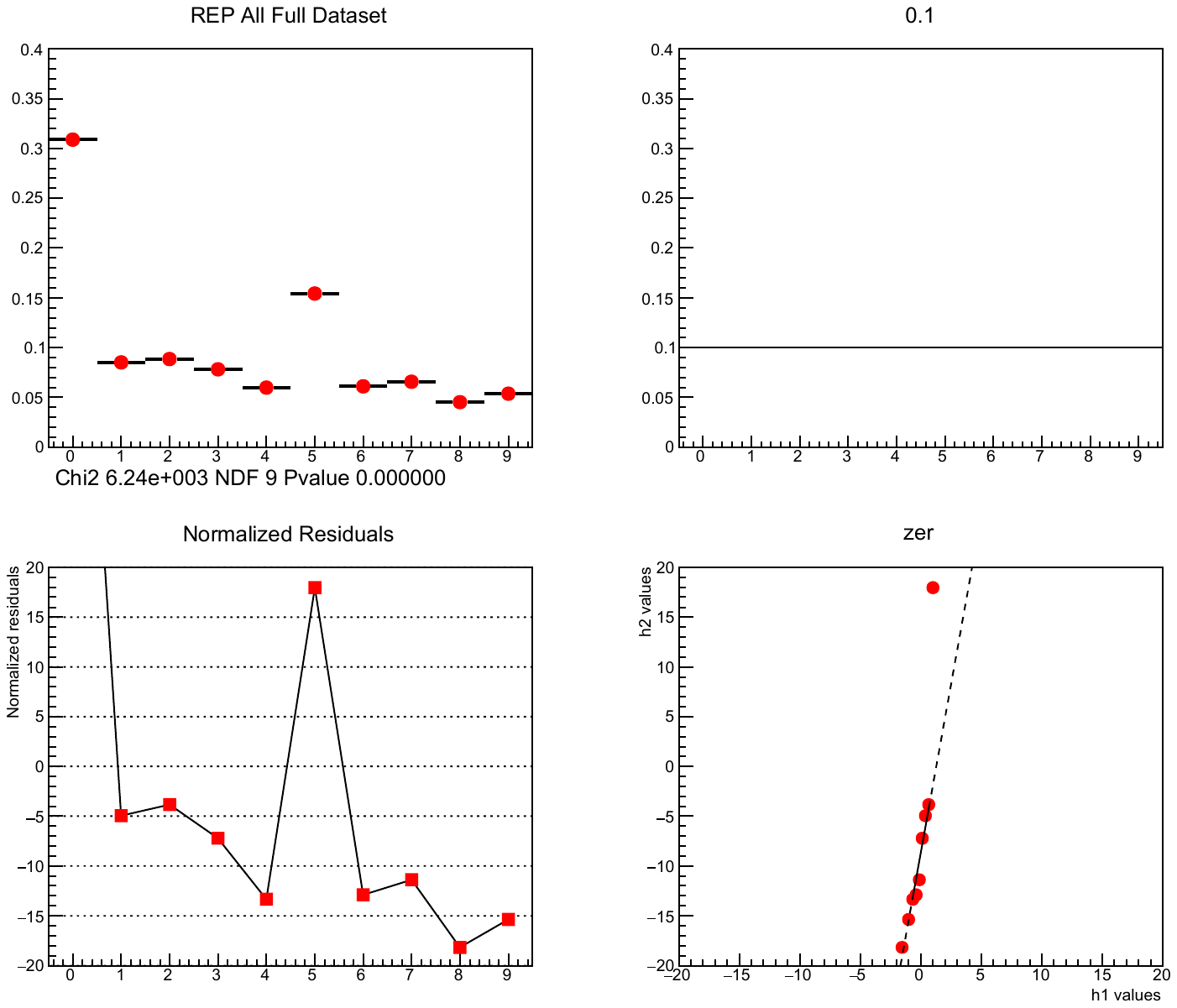}
\includegraphics[width=0.48\textwidth,page=15]{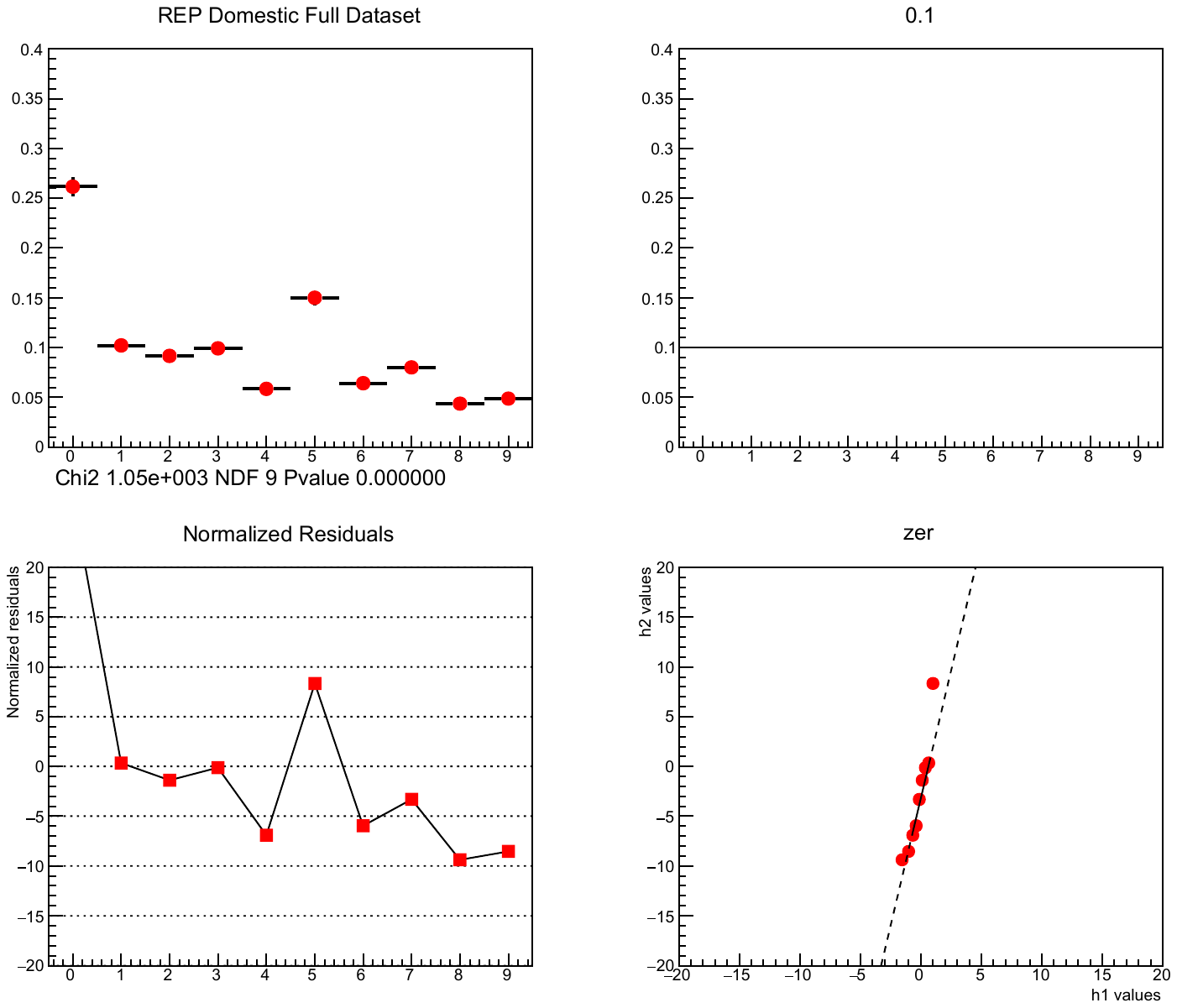}
\includegraphics[width=0.48\textwidth,page=16]{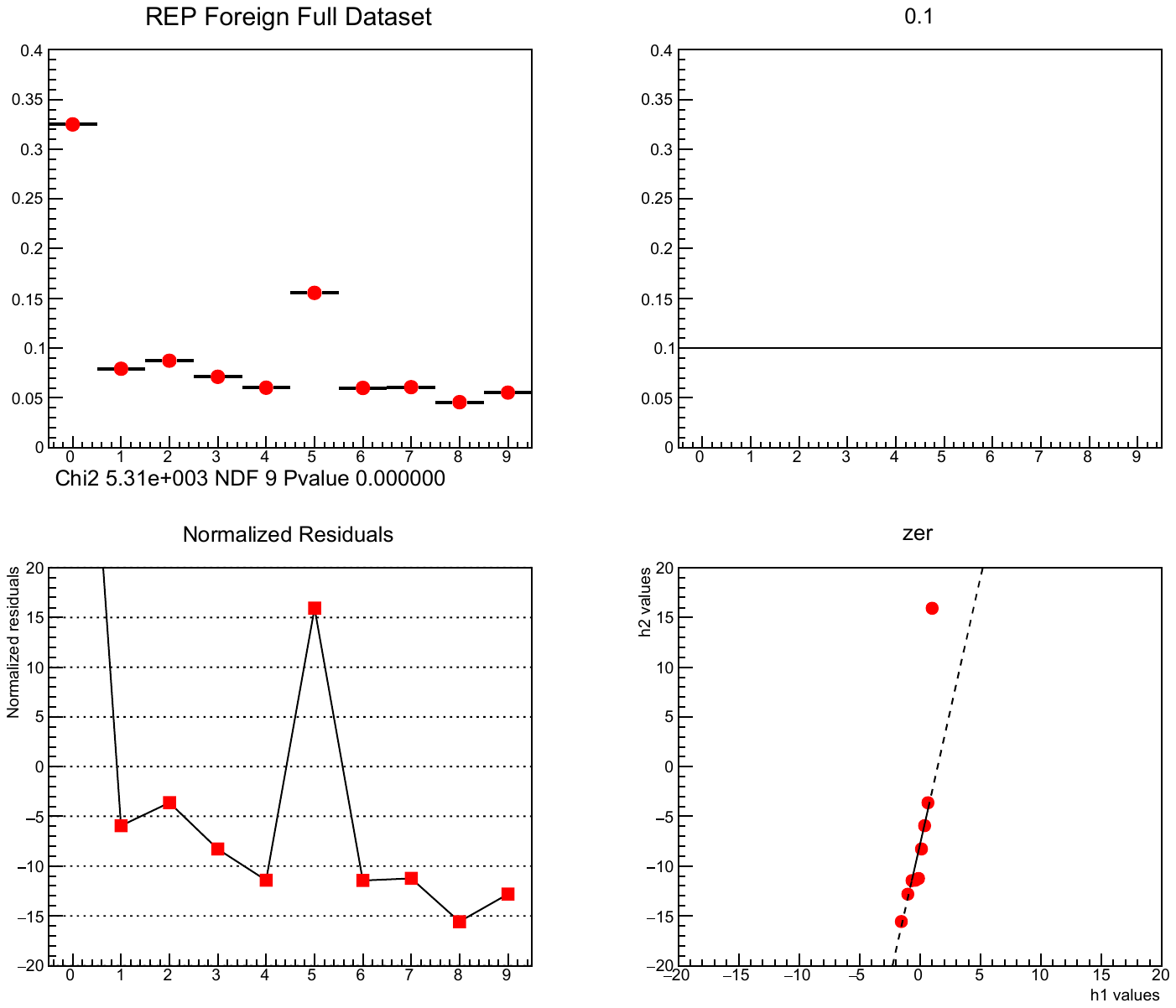}
\caption{$\chi^2$ test comparing various distributions of the least significant digit for $10<k\leq 100$. Top Left and right: the two histograms (or histo and flat distribution function) being compared. Bottom left: Normalized residuals, Bottom right: QQ plot. REP.  
 }
\label{digitsR}
\end{center}
\end{figure}

Figure \ref{digitsS} shows the $\chi^2$ test for STA comparing various  distributions of the least significant digit: domestic 1923-1943 - foreign, domestic 1923-1943 - domestic full dataset, domestic 1923-1943   - flat, domestic full dataset  - flat. From the resulting Chi2 and p value it is possible to see how the distributions of the domestic 1923-1943 are not incompatible with the flat distributions whereas the foreign one in the same period or the domestic full dataset are incompatible with the flat distribution.  Also 1923-1943 domestic and foreign distributions are incompatible.  A similar behaviour is found for CDS (Figure \ref{digitsC}). 
 Figure \ref{digitsR} shows the $\chi^2$ for REP.

See also   Table \ref{tabledigits} for the P test values.

\begin{table}[!ht]
\begin{center}
\includegraphics[width=1.4\textwidth,page=1]{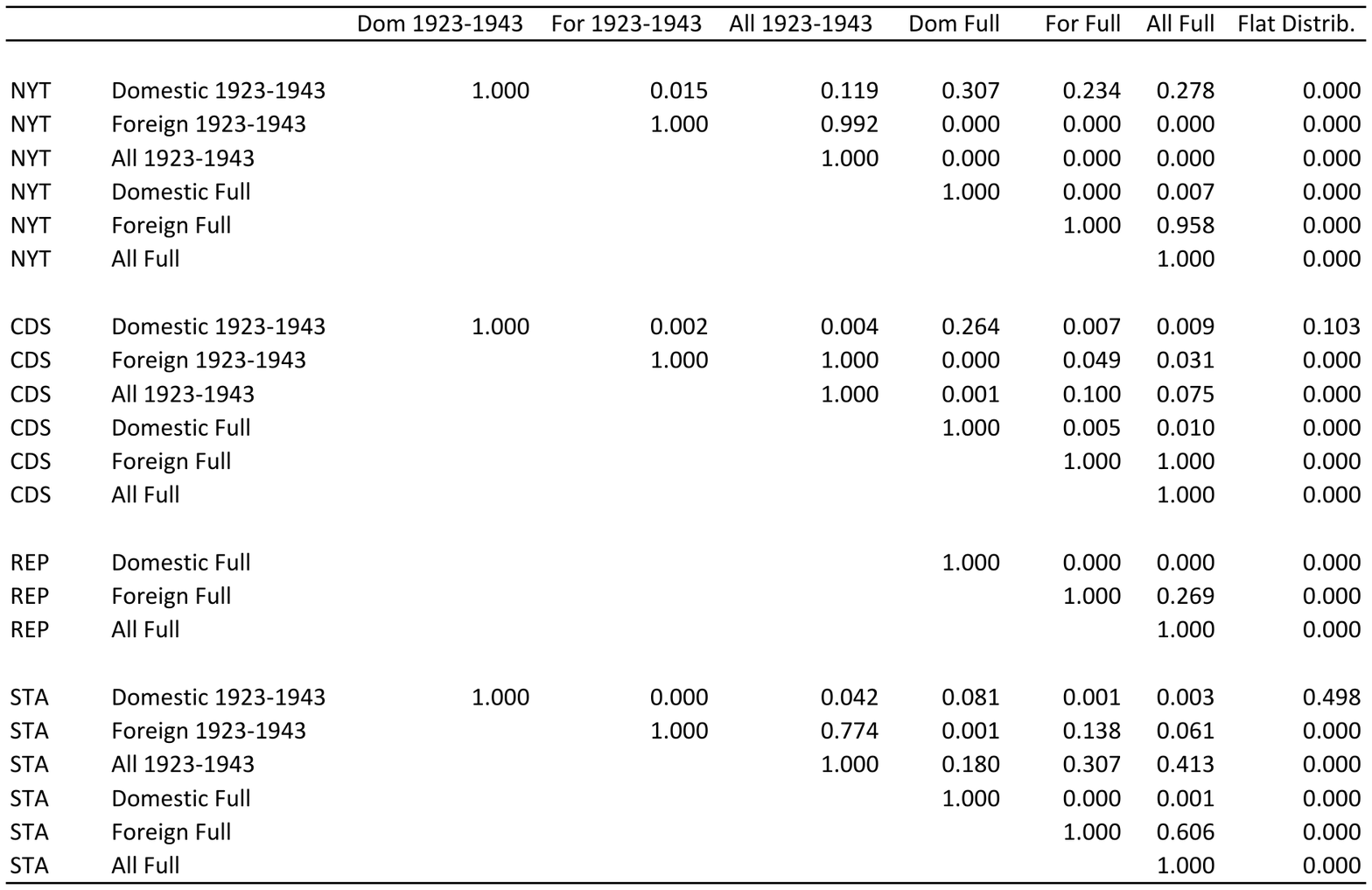}
\vspace{-2cm}
\caption{P value associated to the $\chi^2$ test comparing two histograms    and the flat distribution for each newspaper. We can see that the flat distribution hypothesis is rejected in all cases except for the domestic values of Italian newspapers during fascism.  
 }
\label{tabledigits}
\end{center}
\end{table}

\end{document}